\documentclass[review]{elsarticle}

\usepackage{hyperref}\usepackage{lineno}
\usepackage{multirow}
\usepackage{multicol}
\usepackage{gensymb}
\usepackage{amssymb}
\usepackage{tablefootnote}
\usepackage{caption}
\usepackage{subcaption}
\usepackage{longtable}
\usepackage{tabularx}
\usepackage{amsmath}
\usepackage{mathastext}

\usepackage{wrapfig}
\usepackage{amsmath}
\usepackage[dvipsnames]{xcolor}
\usepackage{comment}

\usepackage{dblfloatfix}
\usepackage{fixltx2e}
\usepackage{placeins}
\usepackage{upgreek}
\modulolinenumbers[1]

\newcommand{\micronsqs}{$\mathrm{\mu m^2}$~}

\newcommand{\microns}{$\mathrm{\mu m~}$}
\newcommand{\micron}{$\mathrm{\mu m}$}

\newcommand{\sxs}{$\mathrm{\sigma_{x}}$~}	

\newcommand{\sts}{$\mathrm{\sigma_{t}}$~}

\newcommand{\nq[1]}{$\mathrm{\cdot 10^{#1} \; n_{eq}/cm^2} $}

\begin{document}

\begin{frontmatter}

\title{First test beam measurement of the 4D resolution of an RSD  450~\microns pitch  pixel matrix connected to a FAST2 ASIC}

\author[1]{L.~Menzio}
\ead{luca.menzio@to.infn.it}
\author[1]{F.~Siviero}
\author[1,2]{R.~Arcidiacono}
\author[1]{N.~Cartiglia}
\author[1,3]{M.~Costa}
\author[1]{T. ~Croci}
\author[1]{M.~Ferrero}
\author[1]{C.~Hanna}
\author[3]{L.~Lanteri}
\author[4]{S.~Mazza}
\author[1,3]{R.~Mulargia}
\author[4]{H-F W. Sadrozinski}
\author[4]{A. Seiden}
\author[1,3]{V.~Sola}
\author[1,3]{R.~White}
\author[4]{M. Wilder}

\address[1]{INFN, Torino, Italy}
\address[2]{Universit\`a del Piemonte Orientale, Italy}
\address[3]{Universit\`a di Torino, Torino, Italy}
\address[4]{University of California at Santa Cruz, CA, US}

\begin{abstract}
This paper reports on the spatial and temporal resolutions of an RSD 450~\microns pitch pixels array measured at the DESY test beam facility. RSDs, Resistive Silicon Detectors, also known as AC-LGAD, achieve excellent position and temporal resolution by exploiting charge sharing among neighboring electrodes.  The RSD matrix used in this study is part of the second FBK RSD production, RSD2, and it is composed of  450~\microns pitch pixel with cross-shaped electrodes.  A 7-pixel matrix was read out by the FAST2 ASIC, a 16-channel amplifier fully custom ASIC  developed by INFN Torino using the 110 nm CMOS technology. The total area covered by the matrix is about 1.5 mm$^2$.  The position resolution reached in this test is  \sxs = 15 \micron, about 4\% of the pitch.  The temporal resolution achieved in this work is \sts = 60 ps, dominated by the FAST2 resolution.   The work also demonstrates that RSD sensors with cross-shaped electrodes achieve 100\% fill factor and homogenous resolutions over the whole matrix surface, making them a suitable choice for 4D tracking applications. 

\end{abstract}

\begin{keyword}
FAST2 \sep Silicon \sep Fast detector \sep Low gain \sep Charge multiplication \sep LGAD \sep 4D tracking
\MSC[2018] XX-XX\sep XX-XX
\end{keyword}

\end{frontmatter}



\section{Introduction}
Silicon sensors based on resistive readout~\cite{8846722}  combine many of the features needed by future experiments: (i) excellent spatial and temporal resolutions, (ii) low material budget (the active part can be a few tens of  \microns thick), (iii) 100\% fill factor, and (iv) good radiation resistance (presently, up to 1-2\nq[15]). In addition, given the large pixel size, RSDs are an enabling technology for constructing 4D silicon trackers~\cite{CARTIGLIA2022167228} with limited power consumption as they reduce the number of readout amplifiers by more than an order of magnitude.  The benefits of resistive readout are maximized when the electrode metal is minimised and shaped to limit the spread of the signal, as reported in a study using a high-precision Transient Current Technique (TCT) setup~\cite{ARCIDIACONO2023168671} to mimic the passage of particles in the sensor.

\section{RSD principles of operation}
\begin{figure}[htb]
\begin{center}
\includegraphics[width=0.95\textwidth]{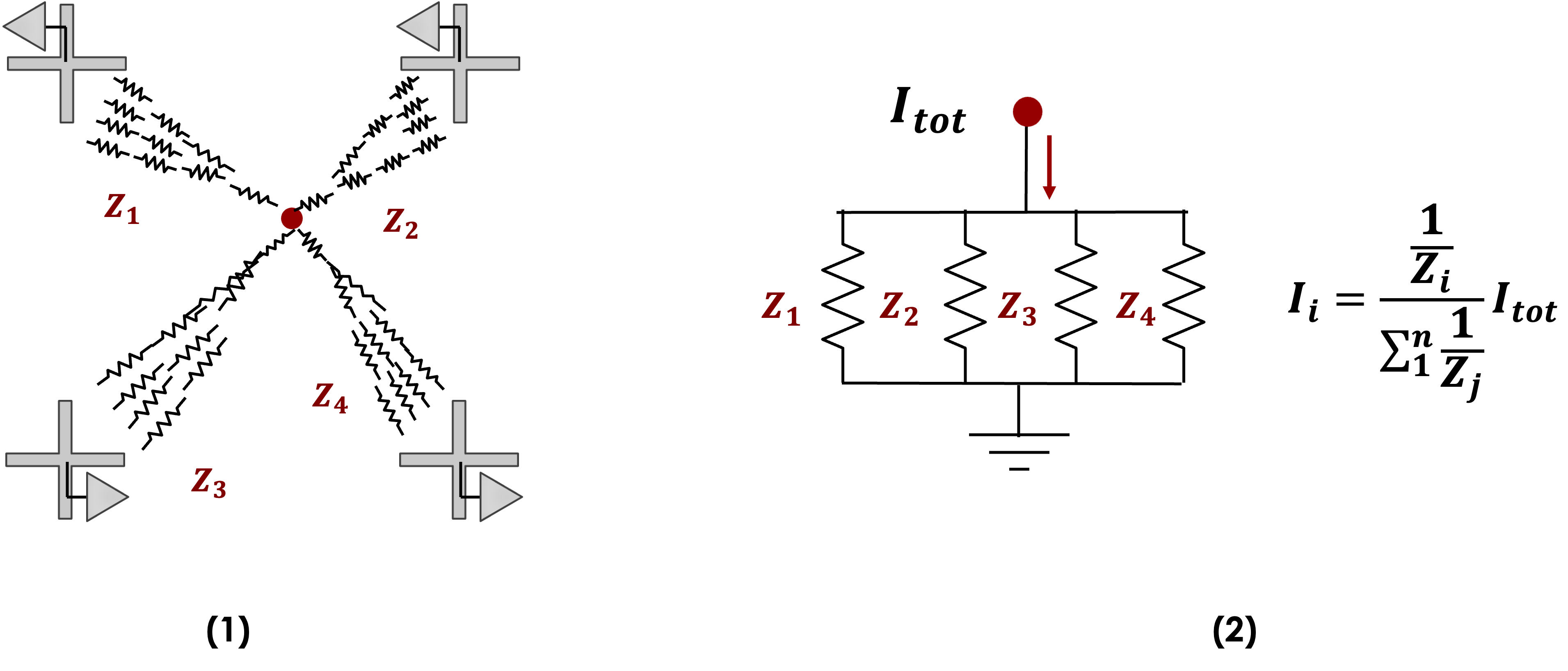}
\caption{Left: the signal splits among the readout electrodes. Right: the split can be computed using the equivalent circuit of a current divider.}
\label{fig:split}
\end{center}
\end{figure}

A short description of the RSD principle of operation is provided in this paragraph; refer to the literature~\cite{ARCIDIACONO2023168671,tornago2020resistive, cartiglia2023resistive} for a complete explanation.
RSDs are thin silicon sensors that combine built-in signal sharing and internal gain. The signal splits among the readout electrodes as a current in an impedance divider, where the impedance is that of the paths connecting the impact point to each electrode, as sketched in Figure~\ref{fig:split} assuming a 4-way split. The input impedance of the front-end electronics must be considerably lower than the path impedances ($Z_{1,2,3,4}$) so that the signal split is governed by  $Z_{1,2,3,4}$.

\section{The DESY test beam facility}
\label{TB_DESY:facility}

The test beam facility in the DESY site at Hamburg-Bahrenfel comprises three distinct beam lines providing electrons or positrons with momenta selectable in the range from 1 up to 6 GeV~\cite{DESY_TB_FACILITY}.  

The test beam campaign reported in this paper was performed in the T24 experimental area.  This area is instrumented with an EUDET2-type telescope~\cite{EUDET} with six planes of 54~\micron-thick MIMOSA-26 Monolithic Active Pixel Sensors~\cite{5402399}.  The EUDET2 performance depends on the six planes' relative positions, the beam energy, and the DUT material budget. With a distance of 38 mm between the planes and a beam momentum of 5.6 GeV/c,  a resolution of $\sigma_{x}$ = 2.89 \microns was achieved.  An EUDET Trigger Logic Unit~\cite{Cussans2009DescriptionOT} provides the trigger to the telescopes. The data acquisition is performed in the EUDAQ framework.

\section{The experimental setup}
In this paragraph, the key elements of the experimental setup are presented. The system comprises the FAST2 ASIC, an RSD2 sensor, a readout board, and the data acquisition system.

\subsection{The FAST2 ASIC}
The FAST2 ASIC~\cite{OLAVE2021164615,9875441}  is designed using standard 110 nm CMOS technology; the power rail is at +1.2 V, and its power consumption is 2.4 mW/ch. It has a footprint of about 5 $\times$ 1.5 mm$^2$. FAST2 has been designed in two versions: (i) an amplifier-comparator version (FAST2D) with 20 readout channels and (ii)  an amplifier-only version (FAST2A) with 16 channels. The  FAST2 front-end circuit comes in two versions, EVO1 and EVO2. Both versions use the same input stage design, a  transimpedance amplifier with two amplification stages, but EVO1 uses standard transistors, while EVO2 uses RF transistors.  The first 8 channels of FAST2A are of the EVO1 type, while the other 8 are of the  EVO2 type. Laboratory tests with a beta telescope have shown that the FAST2A, when coupled with an LGAD pixel with a capacitance of $\sim$3 pF, reaches a resolution of about 50~ps for an LGAD gain above 20.  FAST2A has two programmable features: (i) an internal test-pulse generator used for calibration and (ii)  the preamplifier gain. Depending on the gain selection,  the bandwidth varies between 230 and 665~MHz and the peaking time between 0.49 and 1.2~ns. If not programmed, the FAST2 ASIC signal amplitude is $\sim$10 mV/fC or, equivalently, has a transimpedance of  $\sim 6\; k\ohm$ with a bandwidth of 460~MHz.

 A newer version of the ASIC has been designed to improve the output signal linearity and lower the input amplifier noise. Given the improved signal-to-noise ratio and linearity, this new ASIC,  FAST3~\cite{9908192}, should lower the ASIC contribution to the total temporal resolution to about 10 ps at an input charge above 20 fC.  In this work, FAST2A has been used.

\subsection{The RSD2 sensor}
\begin{figure}[htb]
\begin{center}
\includegraphics[width=0.8\textwidth]{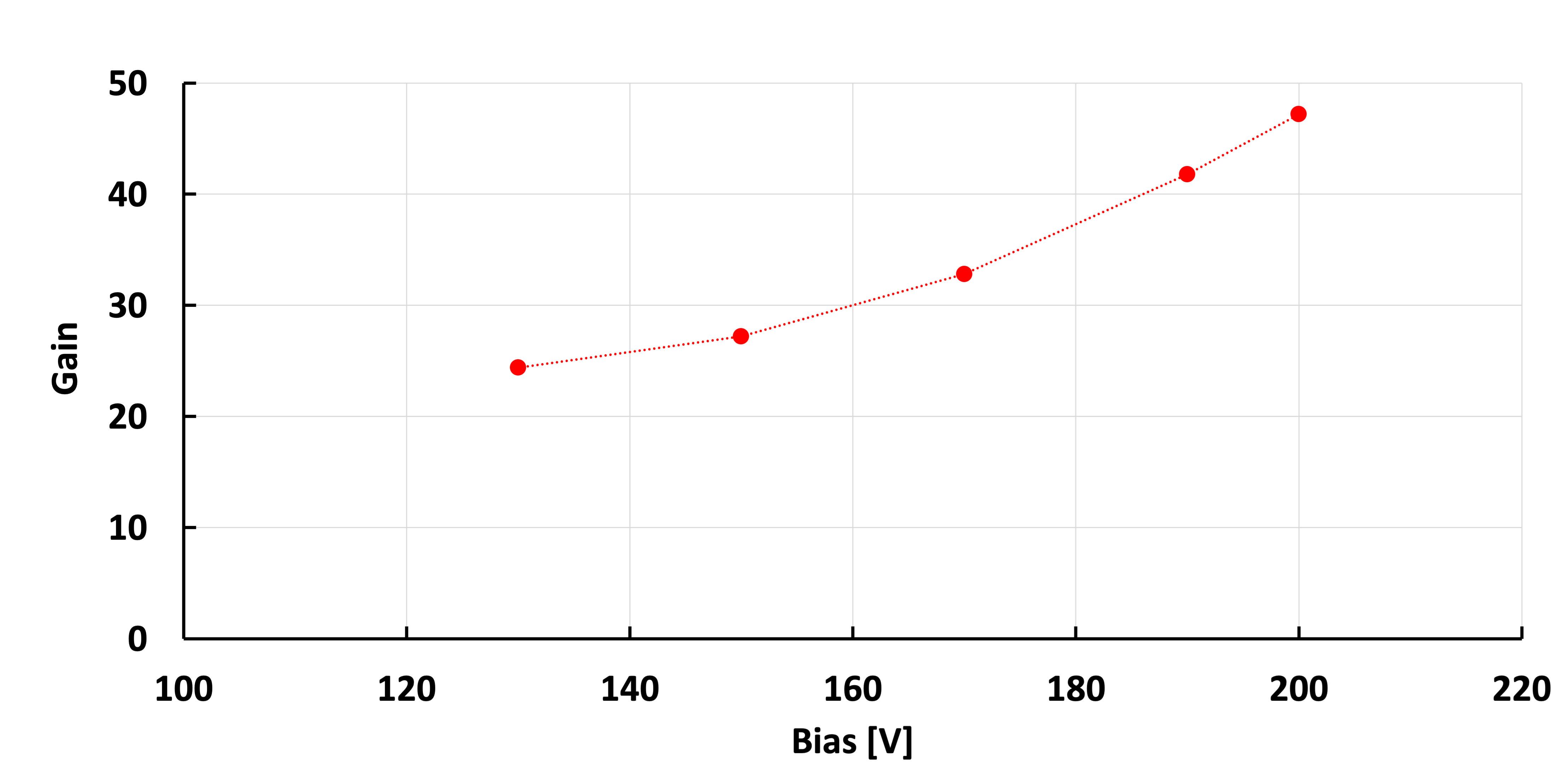}
\caption{The bias versus gain characteristics for the DUT. }
\label{fig:gain}
\end{center}
\end{figure}

The FBK RSD2 production~\cite{Mandurrino_2022} comprises 15 p-type 6" wafers, including epitaxial and float-zone (Si-Si) types.  The active volume is either 45 or 55~\microns thick. The wafers differ in the doping level of the gain implant and the resistivity of the n$^+$ implant. The sensor used in this test is from wafer 4; it has a float-zone 55 \micron-thick active volume. Figure~\ref{fig:gain} shows the gain versus bias curve of the sensor measured at the test beam. The red dots indicate the voltages used in the test described in this paper.  The curve is obtained by converting the FAST2A signal amplitude into charge using the known FAST2 response of $\sim$ 10 mV/fC.

\begin{figure}[htb]
\begin{center}
\includegraphics[width=1.\textwidth]{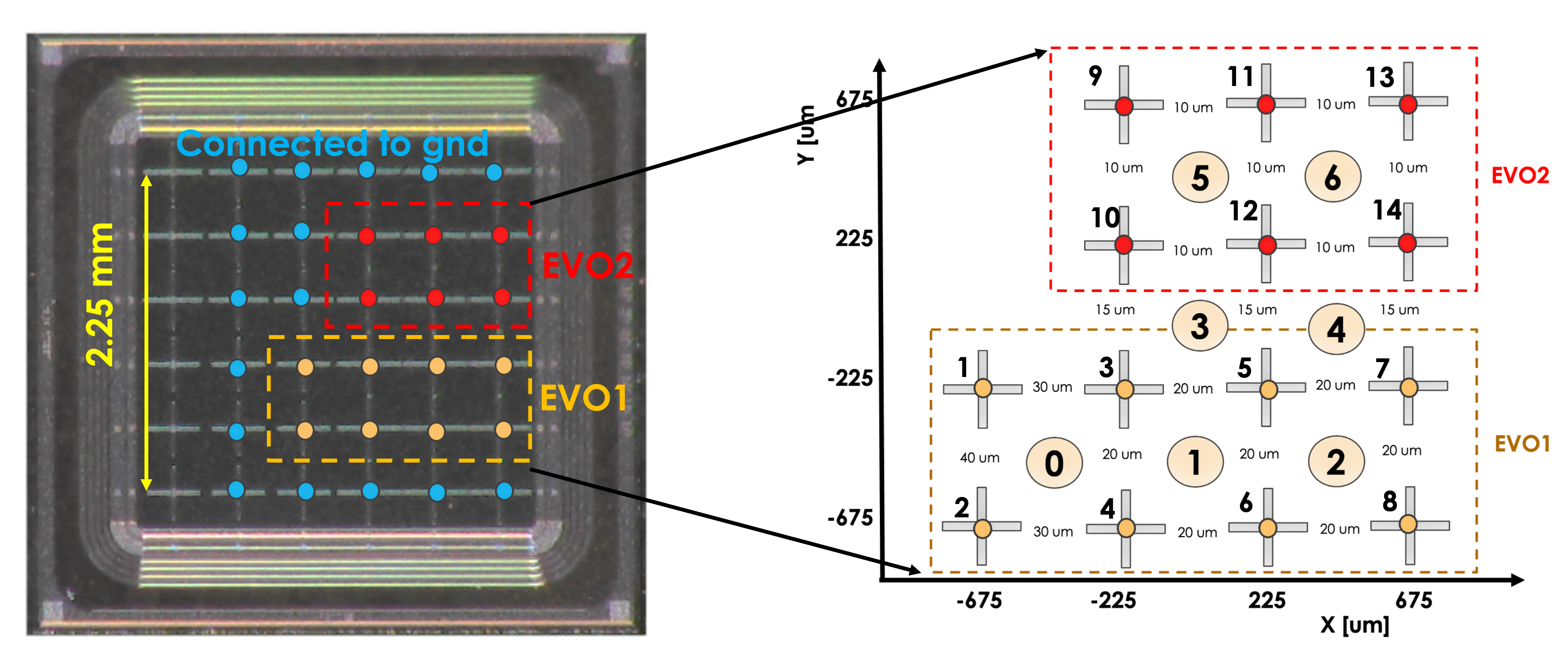}
\caption{Left: a picture of the sensor used in this test, with the electrodes in blue connected to ground, in yellow to EVO1, and in red to EVO2. Right: a schematic of the electrodes, the pixels, the gap between the metal arms, and the x-y reference system used in the analysis. 
}
\label{fig:SR}
\end{center}
\end{figure}

The sensor used in the test is a 6 $\times$ 6 matrix of electrodes with a 450~\microns pitch.  The electrodes are cross shaped, with arms extending in the x and y directions, leaving a small gap between two adjacent arms.  The gap length varies from 10 to 40 \micron, while the width of the arm is fixed at 20 \micron, Figure~\ref{fig:SR}. During the test, 14 electrodes were read out, for a total of 7 pixels; 8 electrodes were connected to the EVO1 channels of FAST2  and the remaining 6 electrodes to  FAST2 EVO2. The other electrodes were all grounded. Figure~\ref{fig:SR} shows on the left side a picture of the sensor, with the electrodes connected to ground in blue, in yellow to EVO1, and in red to EVO2. The right side reports a schematic of the electrodes, the pixels, the gap between the metal arms, and the x-y reference system used in the analysis. 

\subsection{The FAST2-RSD readout board}
\begin{figure}[htb]
\begin{center}
\includegraphics[width=0.9\textwidth]{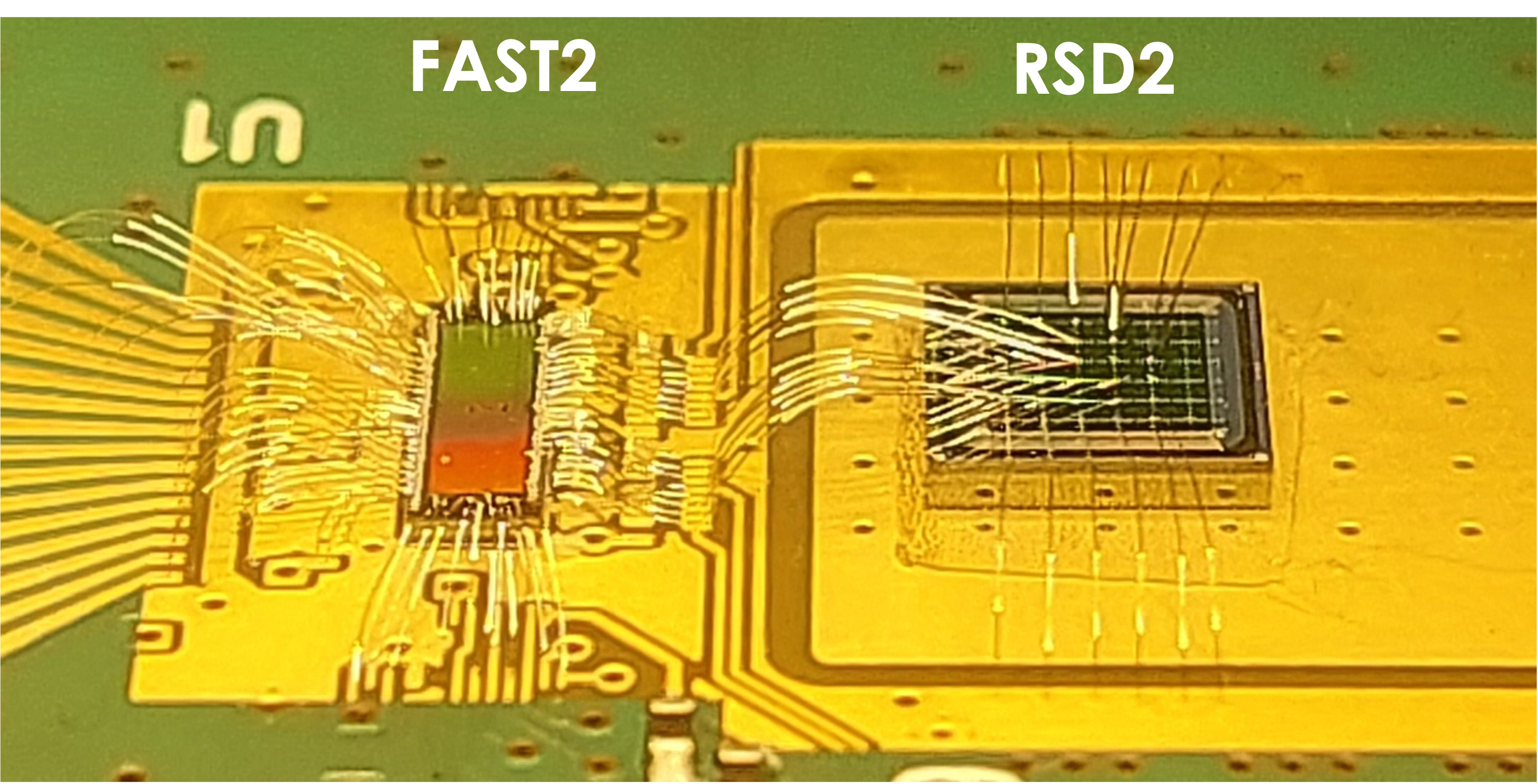}
\caption{Image of the RSD2 sensor wire-bonded to the FAST2 ASIC on the readout board.}
\label{fig:photo}
\end{center}
\end{figure}

The RSD2 sensor and the FAST2 ASIC were mounted on a custom PCB board. The ASIC section of the board is powered by a single voltage line at 4.0 V, and via voltage regulators provides the power to the ASIC. The sensor section of the board provides filtered HV to the sensor. 
The board houses 16 MCX connectors for the FAST2 output signals. Figure~\ref{fig:photo} shows a picture of the board used in this test. The connections from the sensor to the ASIC were made with two sets of wire bonds, using intermediate pads to ensure the possibility of changing the sensor without damaging the ASIC.  The board can be instrumented with an Arduino microcontroller to program the FAST2 settings. The controller was not mounted for this test, so FAST2 worked at the default gain setting. 

\subsection{The acquisition system}
A schematic of the data acquisition system used during the test beam is shown in Figure~\ref{fig:setup}. The setup has a  trigger path (left side of the figure) and a data path (left side of the figure):
\begin{figure}[htb]
\begin{center}
\includegraphics[width=0.95\textwidth]{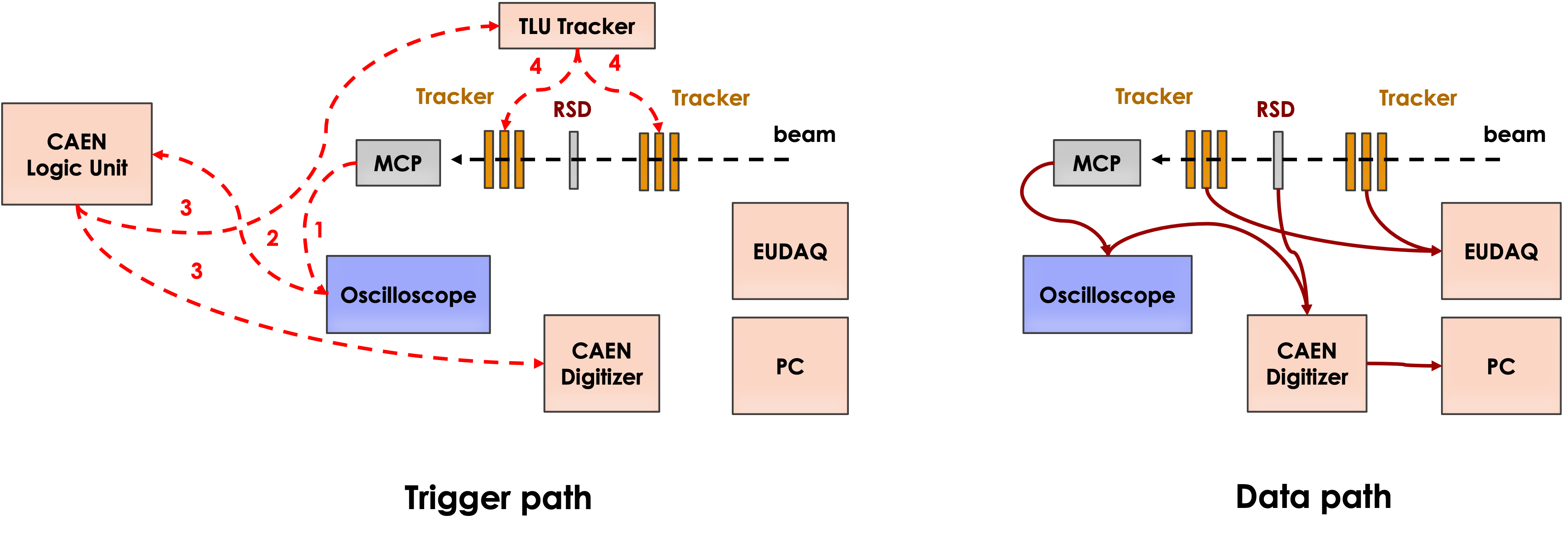}
\caption{Schematic of the data acquisition system.  Left: the trigger path. Right:  the data path.}
\label{fig:setup}
\end{center}
\end{figure}
\begin{itemize}
\item Trigger path: the initial trigger is generated by an electron hitting the Photonis MCP-PMT~\cite{MCP}. The MCP  triggers, in high impedance, (line 1) the LeCroy HD040 oscilloscope~\cite{Teledyne_scope}  that, in turn,  triggers  (line 2) a CAEN logic unit~\cite{CAEN_Logic}. The logic unit triggers (line 3) the AIDA-2020 TLU, provided by the DESY facility,  and the CAEN DT 5742 Digitizer~\cite{CAEN_Digitizer} ($\sim$~500~MHZ analog bandwidth, 5 GS/s) used to read out the DUT. Finally, the TLU triggers (line 4) the EUDET telescope data acquisition. 
\item Data path: upon trigger arrival, the signals from the MCP and the DUT are digitized by the CAEN Digitizer and stored on the DAQ PC, while the data from the telescope are saved on the EUDET PC. The digitizer rate is 5 GS/s, providing 6-7 samples on the signal rising edge. 
\end{itemize}

The bias voltage to the sensor and the MPC is provided by a CAEN DT 1471ET~\cite{CAEN_HV} unit. 
Tracks reconstruction from the telescope data was performed using the Corrivreckan package~\cite{corryvreckan}. It was necessary to employ the General Broken
Lines algorithm to correctly account for the scattering over the telescope and DUT materials of 5 GeV electrons. Noisy pixels in the tracker were masked, and events with multiple tracks were discarded.   The data acquisition systems of the DUT and beam telescope run independently, each producing a file per run, merged offline.  The merging operation checks the possibility that a spurious trigger on either system misaligns the streams of events and, when this happens, realigns the two files.

\begin{figure}[htb]
\begin{center}
\includegraphics[width=1.\textwidth]{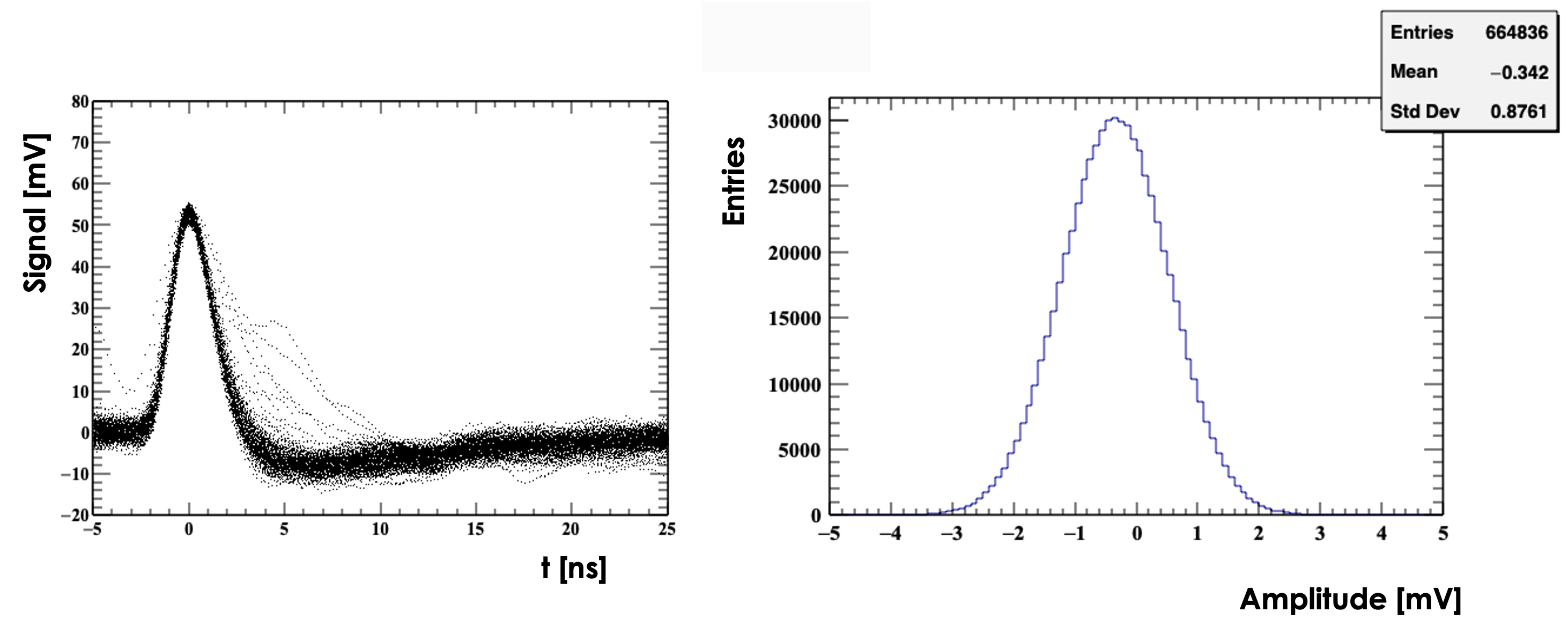}
\caption{Left: FAST2A EVO1 output signals. Right: Signal amplitude of a single sample on the baseline. }
\label{fig:450_Signal}
\end{center}
\end{figure}

Figure~\ref{fig:450_Signal} shows on the left an example of the sensor output signal recorded at the test beam. As expected, the signal is bipolar due to the sensor AC coupling, the signal has a triangular shape determined by the convolution of the RSD output current with the FAST2A and digitizer shaping times (the combined bandwidth is $\sim$~450~MHz); it has a rise time of $\sim$1 ns and a slightly longer fall time.  On the right side of Figure~\ref{fig:450_Signal},  the amplitude of a sample in the signal baseline is reported for all events taken during a given run at the test beam. The amplitude RMS,  $\sim$0.9 mV,  shows that the electronic noise is rather small, yielding a signal-to-noise ratio above 50 for signals above 5 fC (sensor gain of 10).  The amplitude RMS of the sum of two baseline samples separated by 1 ns is 1.37 mV, which is higher than what it would be for fully uncorrelated noise, $0.87 \times \sqrt{2} = 1.23$ mV. The correlation coefficient, calculated as $\mathrm{\rho = \frac{\sigma_{tot}^2 - \sigma_i^2 - \sigma_j^2}{2\sigma_i\sigma_j}}$, is $\rho = 0.24$, and is a measure of common electronic noise and baseline fluctuations.  

\section{Notable quantities used in the analysis}

Table~\ref{tab:def} reports definitions and variables used in this study.

\begin{longtable} {| m{3.cm} | m{9.cm} |}
\hline
 Good events & Events with a track pointing to one of the 7 pixels. \\ \hline
Active electrodes & The 14 electrodes read out during the test. \\ \hline
Active pixels& The 7 pixels read out during the test. \\ \hline
$\sigma^{sample}_i$ & The single-sample amplitude standard deviation of electrode i, measured in absence of signal.  \\ \hline
A$_{i}$ &  The signal amplitude of electrode $i$. The signal amplitude is defined as the peak of the parabola fitted to the  6 highest samples.  \\ \hline
$\sigma^{amplitude}_i$ & The amplitude standard deviation of electrode i. It depends on noise $\sigma^{sample}_i$ and the fitting procedure.  \\ \hline
A$^{max}$ & The highest amplitude among the 14 electrodes. \\ \hline
 A$_{all}$ &  The sum of the 14 amplitudes. This sum is computed when A$^{max}$ is on either electrode 5 or 12 to ensure full signal containment among the active electrodes.  \\ \hline
 A$^{min}_i = 3\times \sigma^{amplitude}_i$ & The minimum detectable  amplitude. \\ \hline
 A$_{pixel}$ &  The amplitude measured by a pixel, defined as the sum of the amplitudes seen by the 4  electrodes.  \\ \hline
A$_{pixel}^{max}$ &  The highest pixel amplitude.  \\ \hline
MPV$_{all}$, MPV$_{pixel}^{max}$, MPV$_{i}$ & The most probable value of the Landau fit to the A$_{all}$, A$_{pixel}^{max}$, A$_{i}$ distributions.  \\ \hline
 $A^{CFD30}_i$ & The 30\% amplitude of the electrode i. The 30\% position is computed using the positions of the sample right above and right below the 30
 \% point.  \\ \hline
$\sigma^{CFD30}_i$ & The standard deviation of  $A^{CFD30}_i$.  It depends on $\sigma^{sample}_i$ and the fitting procedure.  \\ \hline
 $t^{meas}_i$&  The electrode i hit time measured at $A^{CFD30}$.   \\ \hline
 $t^{rise}_i$ & The 0 - 100\% signal rise time of electrode i. \\ \hline
$t^{trigger}$ & The MCP hit time, measured at $A^{CFD30}$.   \\ \hline
\ $\sigma^{trigger}$ & The standard deviation of  $t^{trigger}$, evaluated in the laboratory to be about 12 ps.  \\ \hline

\caption{Definitions and variables  used in this study} 
\label{tab:def}
\end{longtable}

\section{The reconstruction methods}

\subsection{Reconstruction of the hit position}

The determination of the hit position in RSD  is achieved by combining the information from several electrodes, and its resolution can be expressed as:

\begin{equation}
\label{eq:spaceres}
(\sigma^{hit\; pos})^2 =  (\sigma^{pos-jitter})^2 +  (\sigma^{reconstruction})^2 +  (\sigma^{setup})^2 + (\sigma^{sensor})^2.
\end{equation}

\begin{itemize}
\item $\sigma^{pos-jitter}$: for a single electrode i, this term depends linearly on the uncertainty of the amplitude determination $\sigma^{amplitude}_i$  and the signal variation per unit length $dA_i/dx$:  
\begin{equation}
\label{eq:jitterxi}
 \sigma^{pos-jitter}_i = \sigma^{amplitude}_i/(dA_i/dx).
  \end{equation}
 Combining 4 electrodes together and assuming equal noise and amplitude variation with distance for all electrodes, the above expression leads to:
 \begin{equation}
\label{eq:jitterxx}
 \sigma^{pos-jitter}  \propto  \frac{\sigma^{amplitude}}{\Sigma_i A_i}\times pitch.
 \end{equation}

 \item $  \sigma^{reconstruction}$: term that depends on the position reconstruction method
\item $  \sigma^{setup}$: due to hardware-related effects such as differences in gain among amplifiers or misalignment between the device under test and the reference tracking system. 
\item $\sigma^{sensor}$: term grouping all sensor imperfections contributing to an uneven signal sharing among electrodes, for example, non-uniform $n^+$ implant. 
\end{itemize}
The jitter term decreases with the sum of the signal amplitudes, while the other three terms contribute to the constant term, the systematic limit of the measurement.

The hit position was reconstructed using two different algorithms:  (i) the Discretized Position Circuit (DPC)~\cite{507162}, and (ii) the Sharing Template (ST). 

\subsubsection{The Discretized Position Circuit (DPC) reconstruction method}

\begin{figure}[htb]
\begin{center}
\includegraphics[width=1.\textwidth]{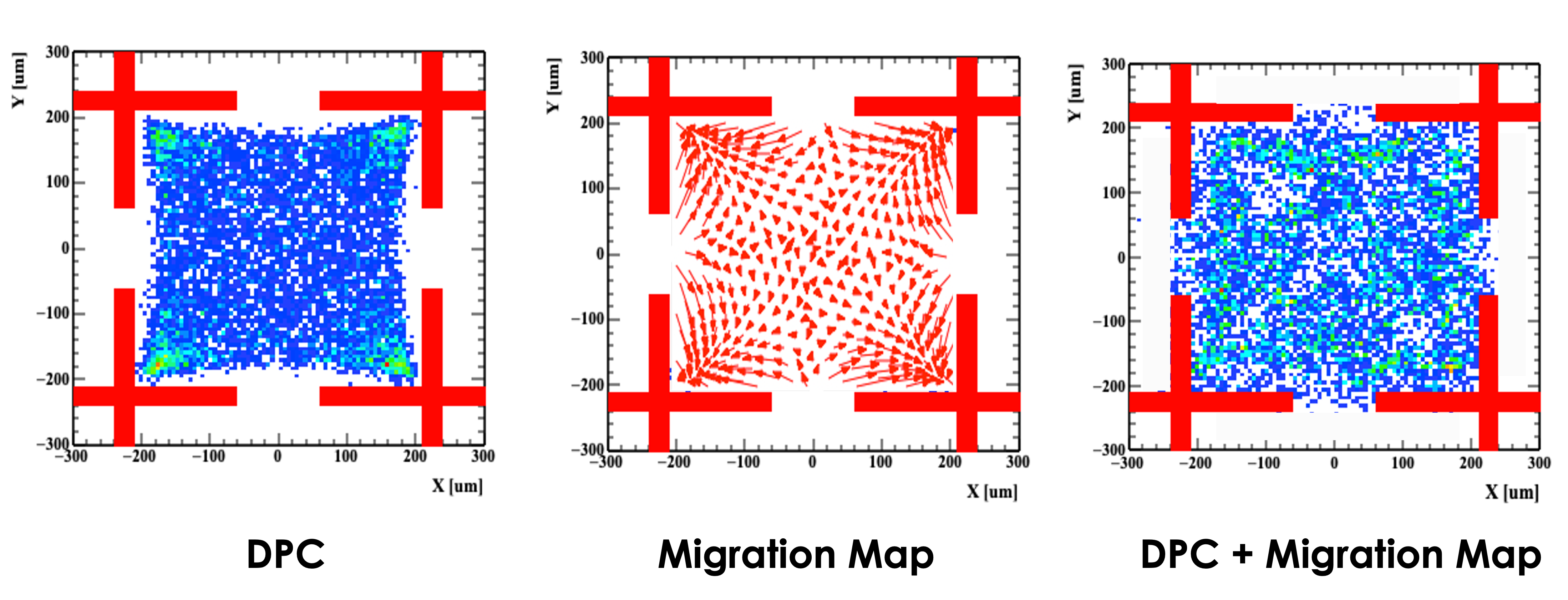}
\caption{Left: hit positions reconstructed using the DPC equation. Middle: migration map. Right: hit positions corrected using the migration map. }
\label{fig:DPC}
\end{center}
\end{figure}

In DPC,   the position is reconstructed using the signal amplitude imbalance between the two sides (right - left, top - bottom) of the pixel. Using as an example pixel 0 of Figure~\ref{fig:SR}, the DPC equations are:
\begin{equation}
\begin{aligned}
\label{eq:dpc}
 x^{meas} = x_0 +  k_x*\frac{ (A_3 + A_4) - (A_1+A_2) }{ \Sigma_1^4 A_i} \\
 y^{meas} = y_0 +  k_y*\frac{ (A_1 + A_3) - (A_2+A_4) }{ \Sigma_1^4 A_i} ,
\end{aligned}
\end{equation}

where $A_i$ is the signal amplitude measured on the electrode $i$, $x_0$ and $y_0$ are the coordinates of the central point of the pixel, and $k_x$ and $k_y$ are given by:

\begin{equation}
\begin{aligned}
\label{eq:kdpc}
k_x = \frac{pixel \; size}{2}* \frac{1}{\frac{ (A_3 + A_4) - (A_1+A_2) }{ A_1+A_2+A_3+A_4} |_{x =x_3}}\\
k_y = \frac{pixel \; size}{2}* \frac{1}{\frac{ (A_1 + A_3) - (A_2+A_4) }{ A_1+A_2+A_3+A_4} |_{y = y_3}}.
\end{aligned}
\end{equation}
For the sensor in this test, the coefficients that lead to the best results are $k_x, k_y = 1$ .  
As explained in ~\cite{ARCIDIACONO2023168671}, the x,y coordinates calculated with equations~\ref{eq:dpc} suffer from systematic shifts; this effect can be seen in the left plot of  Figure~\ref{fig:DPC}. This distortion can be compensated by using a migration map,  shown in the middle plot of Figure~\ref{fig:DPC}.  The plot on the right shows the DPC coordinates after the correction.

The migration map used in this analysis was computed at the TCT laser setup in the Torino laboratory employing a sensor with the same layout but different n$^+$ resistivity mounted on a FNAL 16-channel amplifier board~\cite{apresyan2020measurements}. The procedure was as follows: (i) the laser was shot in well-controlled positions, (ii) the x,y coordinates of the hit were reconstructed using equations~\ref{eq:dpc}, and (iii) the arrows were computed connecting the laser system and reconstructed positions. 
The migration map corrects for the shifts introduced by the reconstruction method, the inhomogeneities of the n$^+$  resistive layer, and differences in the electronics input impedance. Since these last two terms are sub-leading~\cite{ARCIDIACONO2023168671},  the migration map computed for a given sensor can be used for all sensors with the same geometry.

\subsubsection{The Sharing Template (ST) reconstruction method}

\begin{figure}[htb]
\begin{center}
\includegraphics[width=1\textwidth]{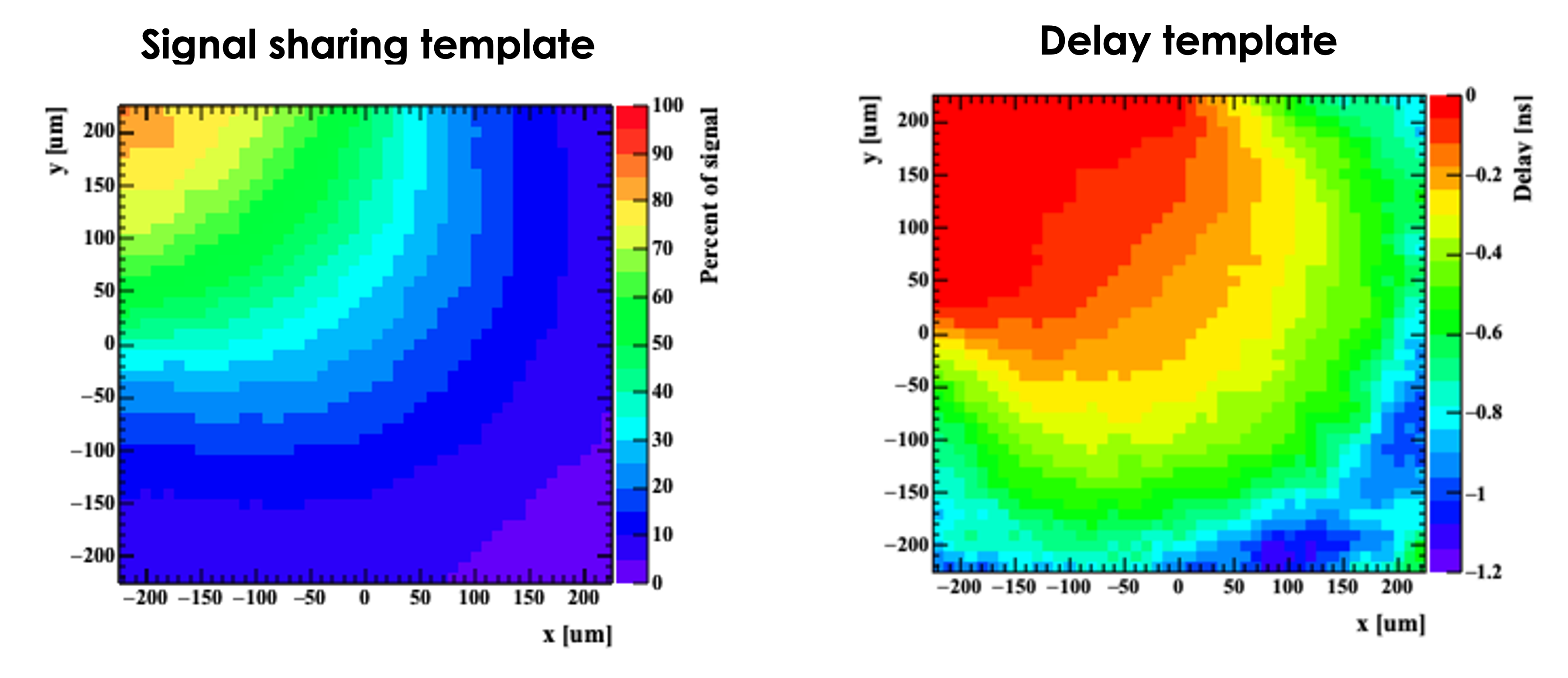}
\caption{Templates for the top left electrode of a pixel: Left:  percentage of the signal as a function of position. Right: signal delay as a function of position.}
\label{fig:ST}
\end{center}
\end{figure}

The second position reconstruction method, ST,  uses a template of the signal sharing among the 4 electrodes as a function of the hit position in the pixel. For each position, the fraction of signal in the 4 electrodes is tabulated in a template.  For every event, the fraction of signal in each electrode is computed and compared with the prediction of the template.  The left plot of Figure~\ref{fig:ST} shows the percentage of the signal seen by the top left electrode of a pixel, as computed using data collected at the DESY test beam. In this study, the template is computed in a grid of 10 $\times $  10 \micronsqs cells, and, to increase statistics,  the events from all 7 active pixels are summed together. The events for the analysis and those used to build the templates belong to different data sets.

The procedure is as follows:
\begin{itemize}
\item For each cell $k$ of the template, the sum of the amplitude-weighted differences between the measured  ($f_{i}^{measured}(k) $) and tabulated ($f_i^{tabulated}(k)$) signal fractions on the 4 electrodes is computed: $\chi^2(k) = \Sigma_{i=1}^4 (( f_{i}^{measured}(k) - f_i^{tabulated}(k))*A_k)^2$, where $i$ is the electrode index. 
\item The coordinates of the cell $k$ with the minimum  $\chi^2(k)$  provide the seed position.
\item The hit position is computed as the $\chi^2$ weighted centroid of the 3$\times$3 cells centered at the seed cell. 
\end{itemize}

\subsection{Reconstruction of the hit time}

\begin{figure}[htb]
\begin{center}
\includegraphics[width=0.6\textwidth]{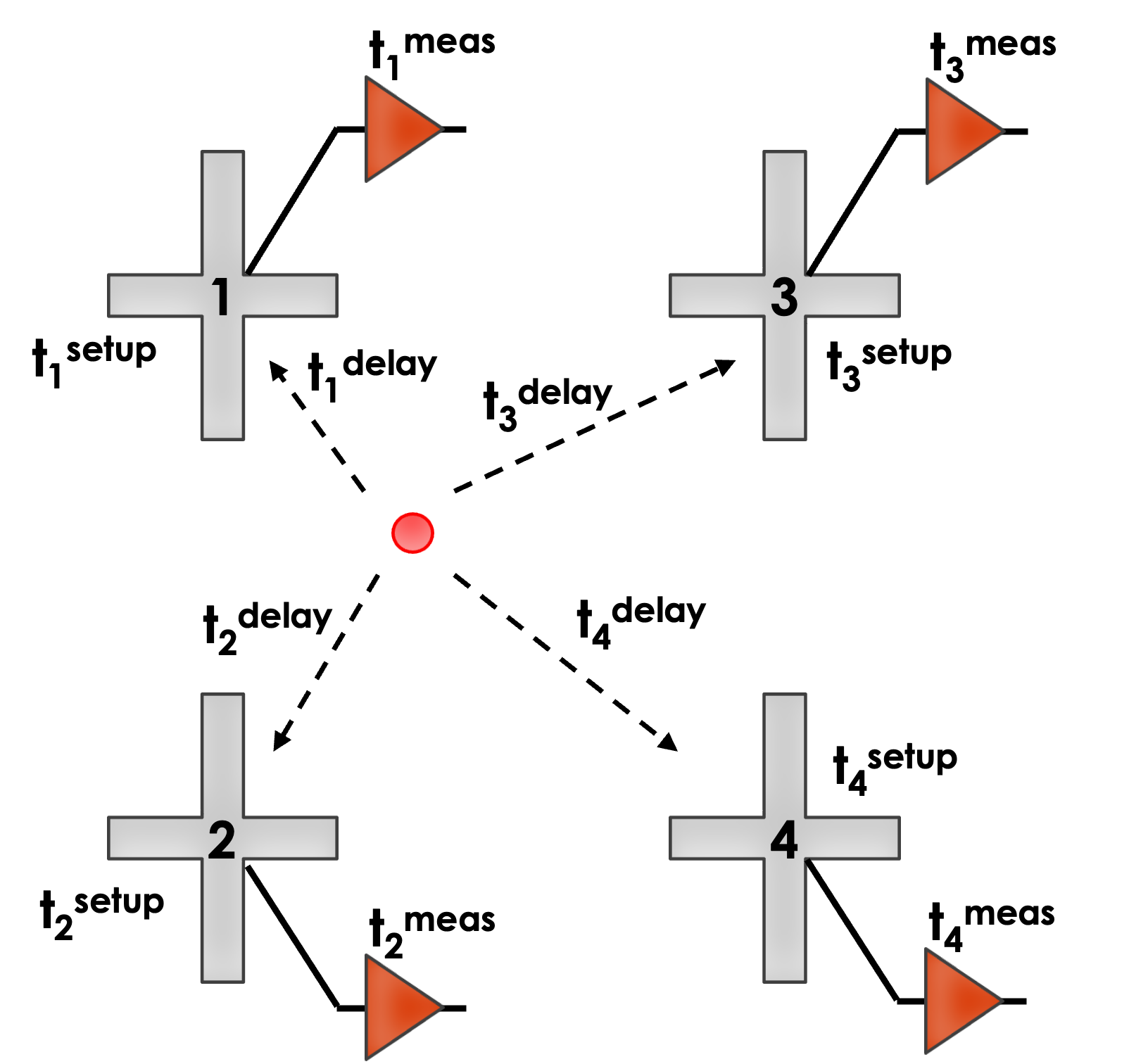}
\caption{Schematic of the various contributions to $t_{i}^{rec}$.}
\label{fig:ttrue}
\end{center}
\end{figure}

In contrast with the hit position reconstruction, where the information from multiple electrodes is needed,  the hit time reconstruction is performed separately by each electrode.  For each electrode, $i$, the measured time, $t_i^{meas}$, differs from the hit time due to the delay, $t_i^{delay}$, introduced by the signal propagation on the resistive layer. 
Therefore, the reconstructed hit time $t_i^{rec}$ can be expressed as: 
\begin{equation}
\label{eq:timetrue}
t_i^{rec} = t_i^{meas} +t_i^{delay}(x,y)+t_i^{setup}
\end{equation}

where $ t_i^{setup}$ is a hardware-specific offset due to PCB traces and cable lengths. The delay as a function of position has been tabulated in a  template computed using test beam data,  shown in the right plot of Figure~\ref{fig:ST}. Given the cross-shaped electrodes, the delay does not increase linearly with distance but has a broad region near the electrode where the signal has a minimum delay.  Figure~\ref{fig:ttrue} illustrates the various contributions to $t_{i}^{rec}$.

The temporal resolution associated with  $t_i^{rec}$ is the sum of 3 terms:

\begin{equation}
\label{eq:timeres}
(\sigma^{rec}_i)^2 =   (\sigma^{time-jitter}_i)^2 + (\sigma^{Landau \; noise}_i)^2 + (\sigma^{delay}_i)^2 
\end{equation}

where:
\begin{itemize}
\item $\sigma^{time-jitter}_i$  depends linearly on the noise $\sigma^{CFD30}_i$ and  the signal derivative at $A^{CFD30}$:
\begin{equation}
\label{eq:jittertt}
\sigma^{time-jitter}_i = \sigma^{CFD30}_i/(dV/dt)|_{A^{CFD30}} \sim \frac{\sigma^{CFD30}_i } { A_i}\times  t^{rise}_i,
\end{equation}
\item $\sigma^{Landau \;noise}_i$ is due to non-uniform ionization. Assuming a 50~\microns thick sensor, this term is about 30 ps.
\item $ \sigma^{delay}_i$ is due to the uncertainty on the hit position reconstruction, it can be minimized with a good determination of the impact point.
\end{itemize}

The uncertainties of the 4 electrodes are not independent since a part of $\sigma^{time-jitter}$ might be due to common electronic noise,  $\sigma^{Landau \;noise}$ is the same for the 4 electrodes (in a given event, the same signal shape is seen by the 4 electrodes),  and $\sigma^{delay}$ affects all $t_i^{rec}$.  The 4 $t_i^{rec}$ estimators can be combined in a  $\chi^2$ function to estimate the hit time $t_{hit}^{rec}$, however under these conditions the covariance matrix $\Omega$ is not diagonal:

\begin{equation}
\begin{aligned}
\label{eq:thit}
\chi^2 (t_{hit}^{rec}) = \sum_{i = 1}^4  \sum_{j = 1}^4 (t_{hit}^{rec} - t_i^{rec}) \Omega^{-1}_{i,j} (t_{hit}^{rec} - t_j^{rec})  \\ 
\frac{\partial\chi^2}{\partial t_{hit}^{rec} }= 0 \rightarrow 
t_{hit}^{rec}= \frac{ \sum_{i,j=1}^{4} t_i^{rec}(\Omega^{-1})_{i,j}}{\sum_{i,j=1}^4 (\Omega^{-1})_{i,j}} \\
 \end{aligned}
\end{equation}

where $\Omega^{-1}$ is the inverse of the covariance matrix. 

If the uncorrelated jitter term is the dominant source of uncertainty,  $\Omega^{-1}$ is diagonal and $t_{hit}^{rec}$   can be expressed as:

\begin{equation}
\begin{aligned}
\label{eq:tjitter}
 t_{hit}^{rec} \sim \frac{\sum_i^4 t_i^{rec}*A^2_i} { \sum_i^4 A^2_i} \\
 \end{aligned}
\end{equation}

where identical  $\sigma^{CFD30}_i$ and $t^{rise}_i$  are used.

\subsection{Determination of the test beam telescope resolution}
\label{sec:track}

The test beam telescope resolution has been evaluated using  the General Broken Lines  (GBL) track resolution calculator tool~\cite{Spannagel}. This program considers the positions of the 6 telescope planes and the material budget of the DUT (10 mm thick PCB board, and 500 \microns of silicon) to estimate the spatial resolution.  Taking into consideration possible plane misalignments and errors in the evaluation of the material budget, the resolution has been measured to be $\sigma_{telescope} = 8 \pm 1.5$~\microns in the x and y directions.

\section{Data taking and selection}

The events recorded at the test beam were triggered by the MCP. The ratio between the MCP and the active pixels areas suggests $\sim$5\% of good events,  however,  since the position of the DUT was shifted with respect to the  beam spot center, the percentage of good events is about 1.8\%. One run, taken at 130 V, has fewer events since the DUT position was shifted in y by $\sim$-1200 \micron. In 2-3 \%  of good events, none of the 14 electrodes has a signal amplitude above $A_{el-min}$. Given the very high efficiency of silicon sensors, the most likely cause of these events is a poor track reconstruction due to noisy pixels. Table~\ref{tab:runs} lists the properties of the runs used in this work.  

\begin{longtable}
{|c|c|c|c|c|c|}
\hline
Bias [V] &  MCP & Good  & MPV$_{all}$ &  MPV$_{pixel}^{max}$  & Gain  \\ 
             &  Triggers [k]                  &events [k] & [mV] & [mV] &  \\ \hline \hline
130 & 401 &  6.4& 122 & 77  & 24 \\ \hline
150 & 440 &8.3 & 136 & 93 & 28\\ \hline
170 & 480 & 8.9 & 164 & 118 & 33 \\ \hline
190 & 475 & 8.5& 209 &157 &  42 \\ \hline
200 & 665 & 11.1 & 236 & 175  & 47 \\ \hline  \hline
\caption{List of runs used in this work}
\label{tab:runs}
\end{longtable}

The event selection is based on three requirements: 
\begin{itemize}
\item A track pointing to any of the 7 pixels
\item A signal on any of the active electrodes above A$_{el-min}~>=~6.5~mV$
\item $t_i^{rec}$ of the highest signal within 1 ns of $t_{trigger}$.
\end{itemize}

\begin{figure}[htb]
\begin{center}
\includegraphics[width=1\textwidth]{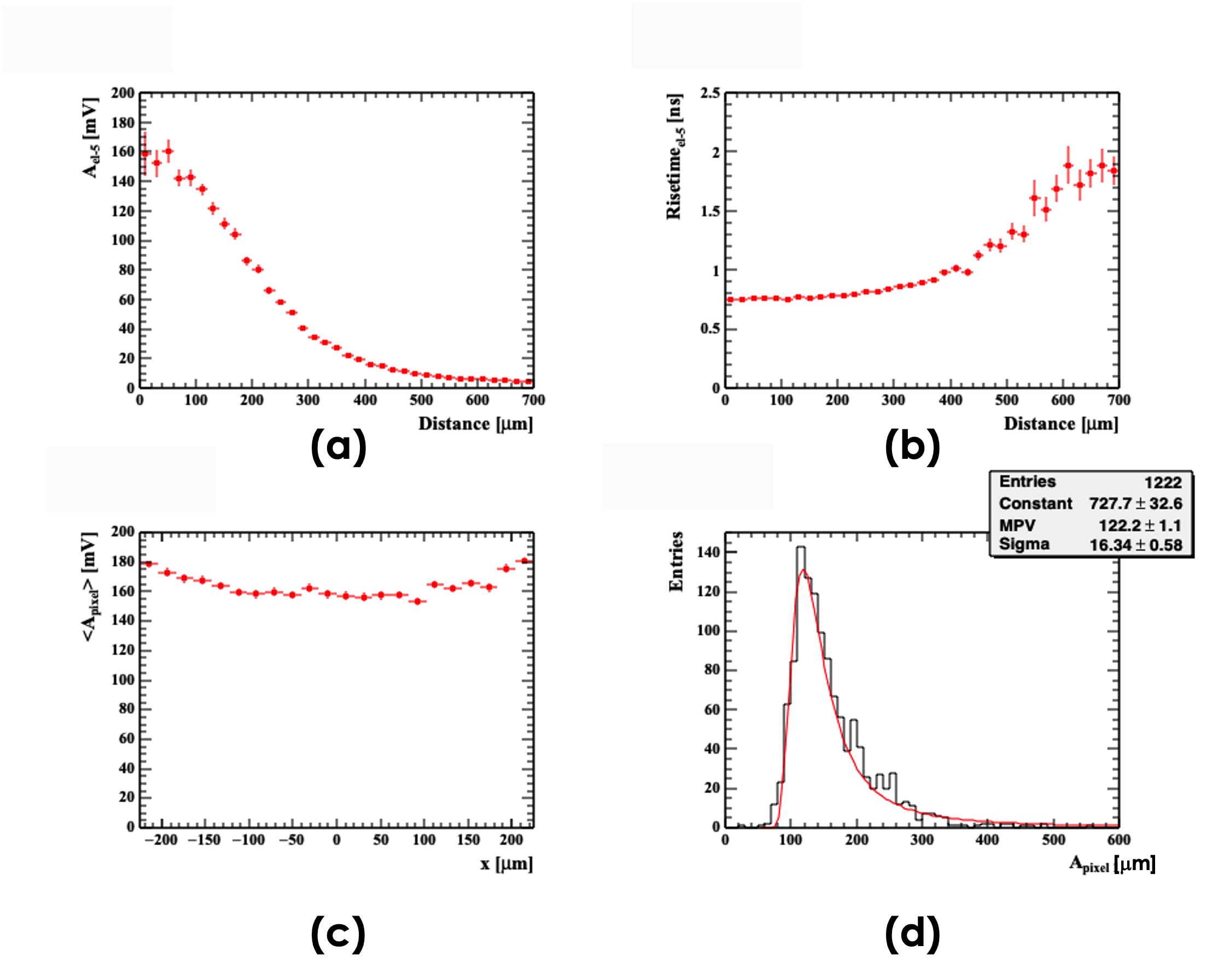}
\caption{(a) Signal amplitude as a function of the hit distance from the electrode. (b) Signal rise time as a function of distance. (c) Projection on the x axis of A$_{pixel}$ over the pixel area. (d) A$_{pixel}$ distribution for all the events in a pixel.}
\label{fig:prop_I}
\end{center}
\end{figure}

Figure~\ref{fig:prop_I} illustrates a few important properties of the events recorded by the DUT:
\begin{enumerate}
\item The signal amplitude on a given electrode (el = 5 in the plot) is visible for a distance of about  $\sqrt{2}~\times~$pitch $\sim$675 \micron, i.e. one pixel. 
\item The signal rise time remains constant for about the arm length, 200 \micron, and then increases with distance. 
\item  The mean value of A$_{pixel}$ is constant at the center of the pixel, increasing by $\sim$10\% at the edge (here, the projection on the x axis is shown). 
\item A$_{pixel}$ follows a Landau distribution as a standard LGAD
\end{enumerate}

\section{Calibration and alignment}

\subsection{FAST2 calibration and saturation}

To ensure uniform response across the 7 pixels,  the amplifier response of the 14 FAST2 channels, 8 EVO1 and 6 EVO2, have been equalized by imposing that the MPV$_i$ is the same for all electrodes.  The calibration constants are the same for all runs and vary between 0.9 and 1.1.  

The FAST2 output signal saturates at about 300 mV ($\sim$ 30 fC) for EVO1 channels and at about 250~mV   ($\sim$ 25 fC) for EVO2 channels. Since saturated signals introduce very strong distortions in the spatial and temporal reconstructions, this analysis did not use events with saturated amplifiers.

\subsection{DUT-telescope alignment}

The telescope and the DUT have been aligned with a software procedure by introducing an x- and y-offset and a rotation. The offset and the rotation were applied to the telescope reference system. 

 The offset was computed by exploiting the fact that the mean value of the telescope x (y) hit positions should be centered on the nominal x (y)  position of the electrode with the highest amplitude. The data were divided into 14 histograms, one per electrode, each containing the telescope hit position for the events where that specific electrode has the highest signal.  The difference between the telescope mean value and the electrode coordinate, $\Delta x_i =  (<x_{track}>~-~x_{el-max = i})$, is the optimal shift for electrode $i$.   
 Given the presence of a rotation, it is impossible to find a single shift in x (y) that minimizes the 14 $\Delta x_i$ at once; what should be minimized is the sum all $\Delta x_i$.
//  $\Sigma_{i=1}^{14} (<x_{track}>~-~x_{el-max = i})$. 

\begin{figure}[htb]
\begin{center}
\includegraphics[width=0.8\textwidth]{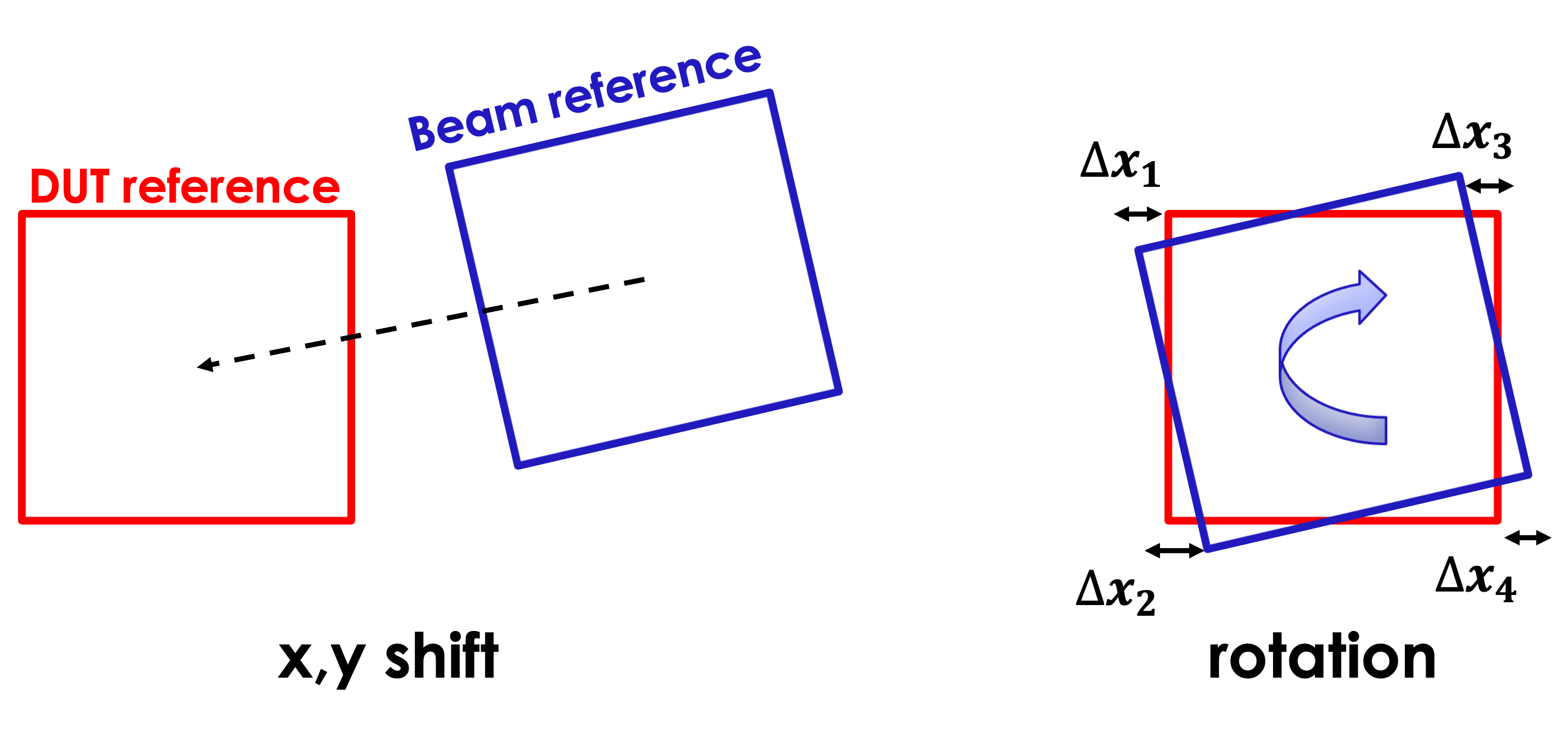}
\caption{Alignment procedure: first, the x-, y- offset is computed, then a rotation is applied. }
\label{fig:align}
\end{center}
\end{figure}

After having determined the best shift, the rotation was computed by minimizing the sum of the absolute value of the distances. // $\Sigma_{i=1}^{14} |<x_{track}> - x_{el-max = i}|$. 
The rotation was found to be  $\theta$ = -0.025 radiants. This procedure is shown in Figure~\ref{fig:align}.
 
\section{Test beam results}

\subsection{Spatial resolution}
\label{sec:sp}
\begin{figure}[htb]
\begin{center}
\includegraphics[width=1\textwidth]{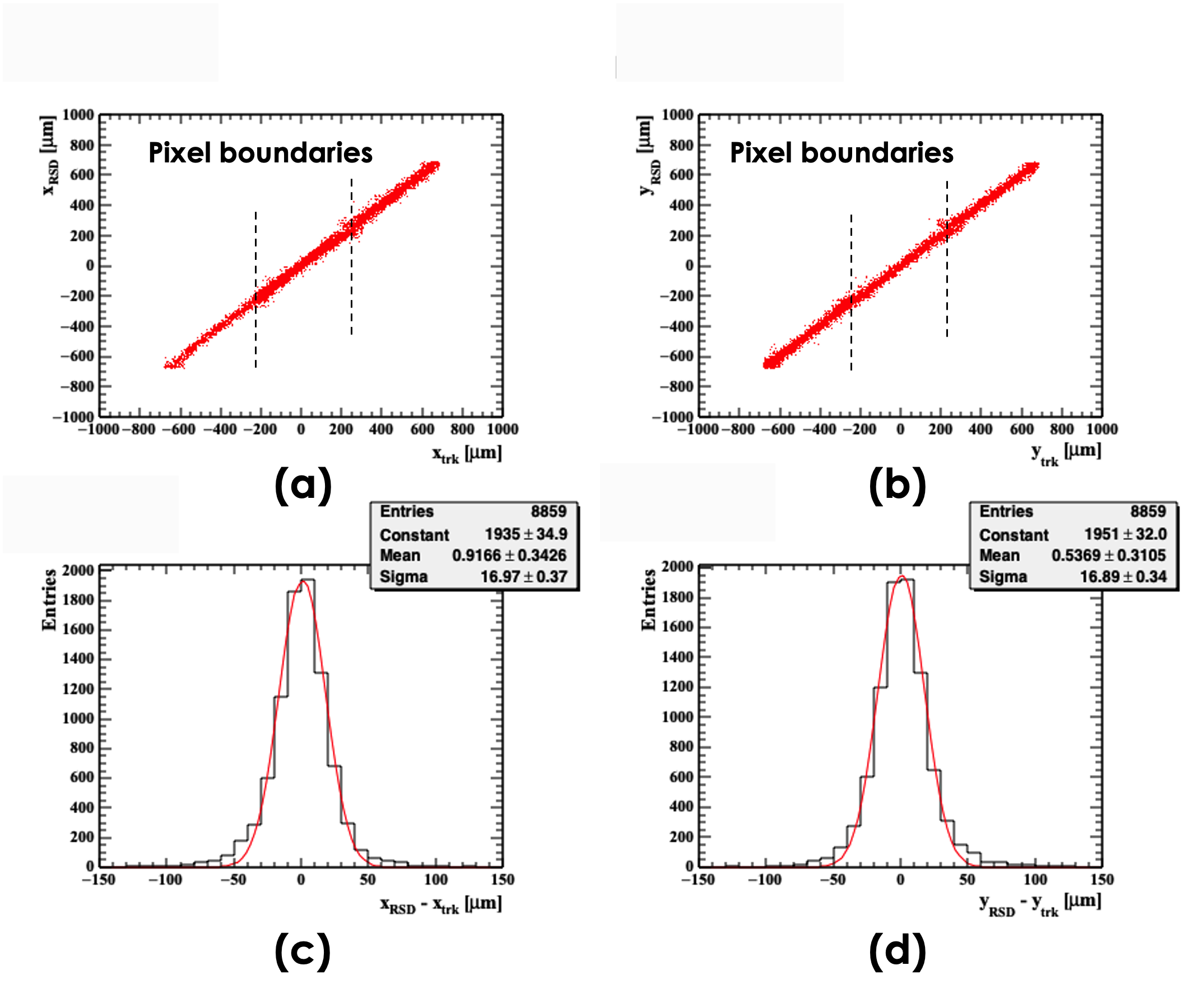}
\caption{(a) and (b): correlation between the tracker and RSD coordinates integrating over all active pixels. (c) and (d): distributions of the x and y tracker and RSD coordinate differences integrating over all active pixels. All plots are obtained using the ST$^{ampl}$ method. }
\label{fig:xres170}
\end{center}
\end{figure}

For each of the bias voltages listed in Table~\ref{tab:runs}, the spatial resolution in x,  y were computed with three different algorithms: (i) the DPC method with the signal amplitude, DPC$^{ampl}$, (ii) the ST method with the signal amplitude, ST$^{ampl}$, and (iii)  the ST method with the signal area, ST$^{area}$, where the area is computed without the signal undershoot.  Figure~\ref{fig:xres170} shows the x-, y- resolution measurements at the bias voltage 170 V using the ST$^{ampl}$ method. The plots (a) and (b)  demonstrate the very good correlation between the telescope hit positions x$_{trk}$,  y$_{trk}$ and the RSD positions, x$_{RSD}$ and y$_{RSD}$, respectively.  Notably, the tracker-RSD excellent correlation continues seamlessly across pixel boundaries, demonstrating that RSDs have 100\% fill factor. 
Plots (c) and (d) report the distributions of the differences (x$_{RSD}$ - x$_{trk}$) and (y$_{RSD}$ - y$_{trk}$) fitted to a Gaussian distribution. The reported values of $\sigma_x = 16.97\;  \mu$m and  $\sigma_y = 16.89\; \mu$m are the convolution of the RSD and telescope resolutions. 
The non-Gaussian tails, defined as differences between the histogram and the fitted distribution in the regions above and below two standard deviations, account for 7.3\% of the data.  The origin of these tails is further discussed in Section~\ref{sec:sp}.

\begin{figure}[htb]
\begin{center}
\includegraphics[width=0.8\textwidth]{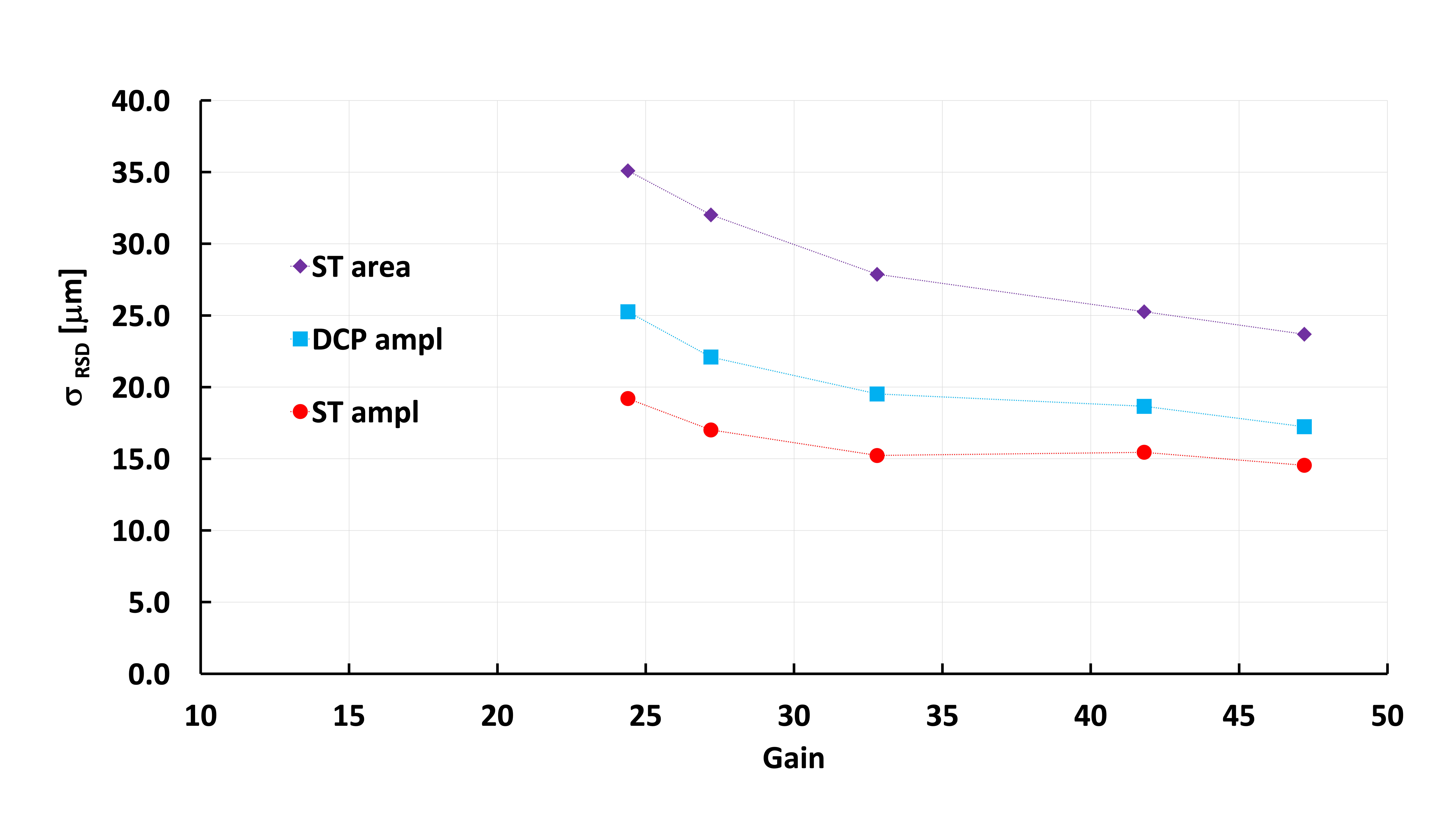}
\caption{RSD resolution for the three reconstruction methods integrating over all active pixels.}
\label{fig:xresall}
\end{center}
\end{figure}

Figure~\ref{fig:xresall}  reports the RSD resolution at each bias voltage for the three methods after the subtraction in quadrature of $\sigma_{telescope}$~=~8~\micron. The results clearly disfavor the choice of ST$^{area}$: as explained in~\cite{ARCIDIACONO2023168671}, the signal amplitude carries more information than the signal area, yielding a better resolution. Both the DPC$^{ampl}$ and ST$^{ampl}$ methods yield very good results. The slightly worse results of the DPC$^{ampl}$ can be understood considering that the migration map was computed using the laboratory TCT setup with a different sensor and electronics while the sharing template of ST$^{ampl}$ was calculated with test beam data and the same hardware.  

The resolution is below 25 \microns for both methods even at the lowest gain, and for ST$^{ampl}$ reaches a constant value of $\sigma_{RSD} \sim$15 \microns for a gain above 30.

The spatial resolution can be fitted as the quadrature sum of a constant and a jitter term, according to Eq.~\ref{eq:spaceres}, given by:
\begin{equation}
 \sigma_{RSD} = \sqrt{(\sigma^{space-constant})^2+ (\frac{\sigma^{amplitude}\times pitch} {\Sigma_i^4 A_i})^2}.
  \end{equation}

\begin{figure}[htb]
\begin{center}
\includegraphics[width=1.\textwidth]{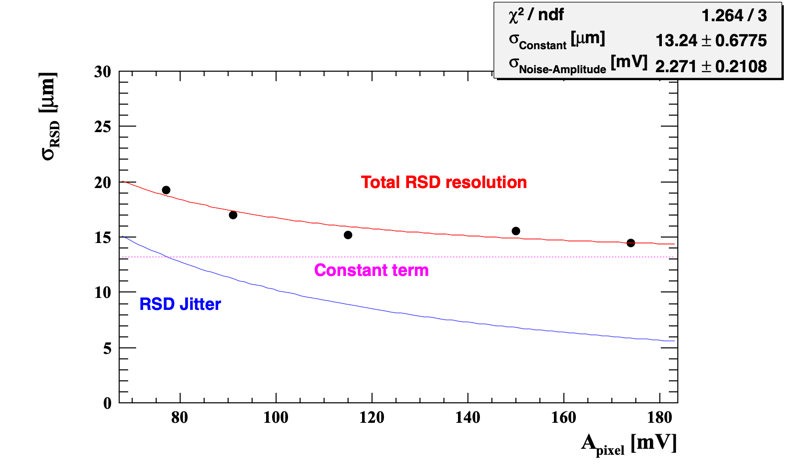}
\caption{Fit to the RSD resolution (ST$^{ampl}$). The constant term dominates the resolution. The points are obtained using the ST$^{ampl}$ method.}
\label{fig:xfit}
\end{center}
\end{figure}

 Figure~\ref{fig:xfit} shows the fit to the ST$^{ampl}$ results as a function of $A_{pixel}$. As expected, the jitter term becomes subleading as the gain increases. Starting at gain $\sim$25, the resolution is dominated by the constant term,  $\sigma^{space-constant} $ = 13.24 \micron. Using the approximation proposed in Eq.~\ref{eq:jitterxx}, the uncertainty on the amplitude determination is computed to be $\sigma^{amplitude} = 2.27 $ mV, about three times $\sigma^{sample}$.   

\begin{figure}[htb]
\begin{center}
\includegraphics[width=1\textwidth]{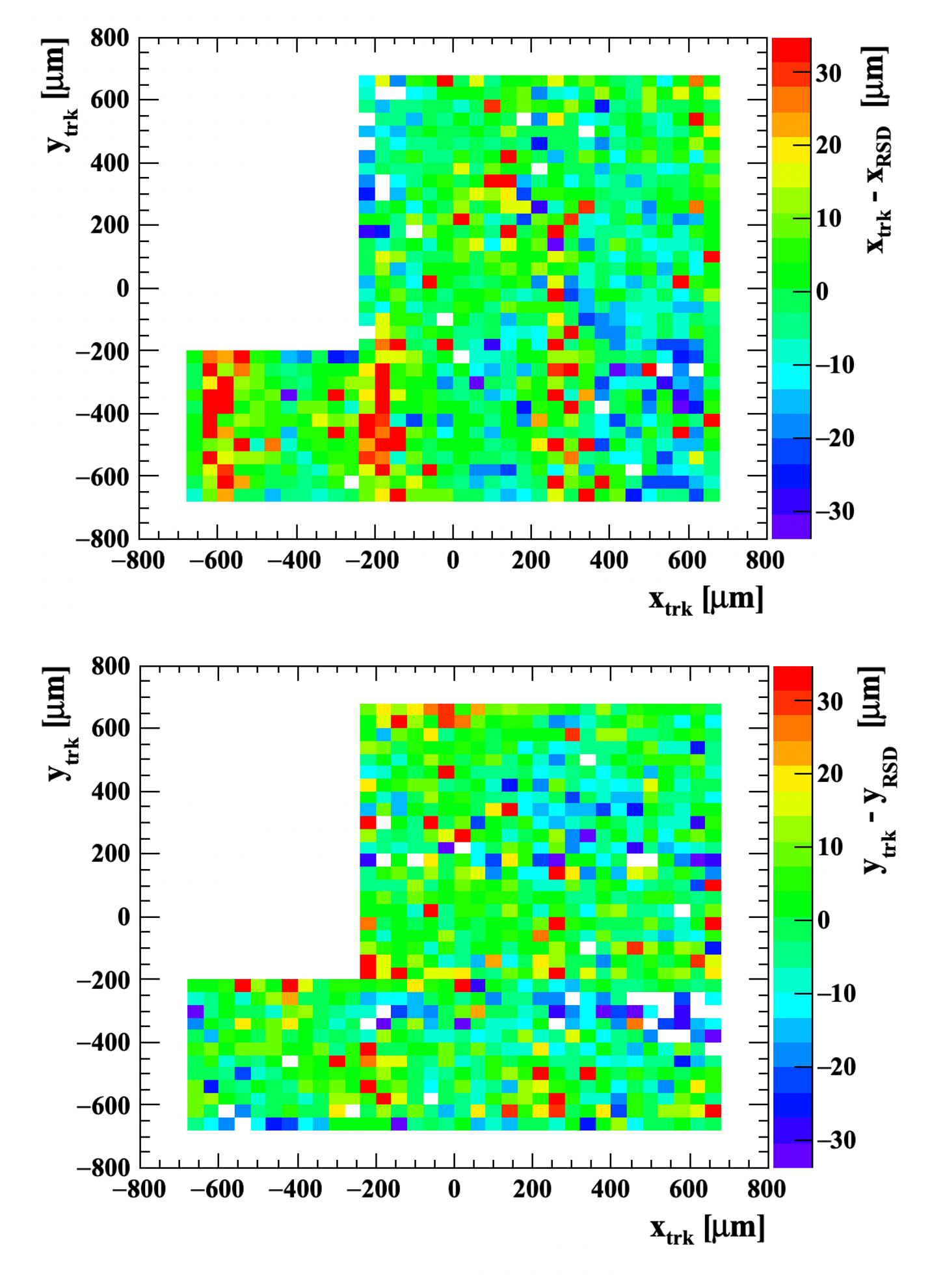}
\caption{Left: the difference x$_{trk}$ - x$_{RSD}$ in the  x$_{trk}$-y$_{trk}$ plane for the 7 pixels used at the test beam. Right: the same plot for the y coordinate. The points below two standard deviations are shown in blue, while those above two standard deviations are shown in red. The plots are obtained using the ST$^{ampl}$ method}
\label{fig:unif}
\end{center}
\end{figure}

The differences  x$_{trk}$ - x$_{RSD}$  and (y$_{trk}$ - y$_{RSD}$)  in the x$_{trk}$-y$_{trk}$  plane are shown for the ST$^{ampl}$ method in Figure~\ref{fig:unif}(left) and (right), respectively, under a bias voltage of 170 V, and the z scale limited to two standard deviations ($\sim$32 \micron).
 With this setting, the points below two standard deviations are shown in blue, while those above two standard deviations are shown in red.  As expected,  the red and blue points cluster around the pixel edges, indicating that these areas have the worst resolution and are the source of the non-Gaussian tails present in  Figure~\ref{fig:xres170}.  

\subsection{Effect of electronic noise or amplifier gain miscalibration on the hit position determination}

\begin{figure}[htb]
\begin{center}
\includegraphics[width=0.8\textwidth]{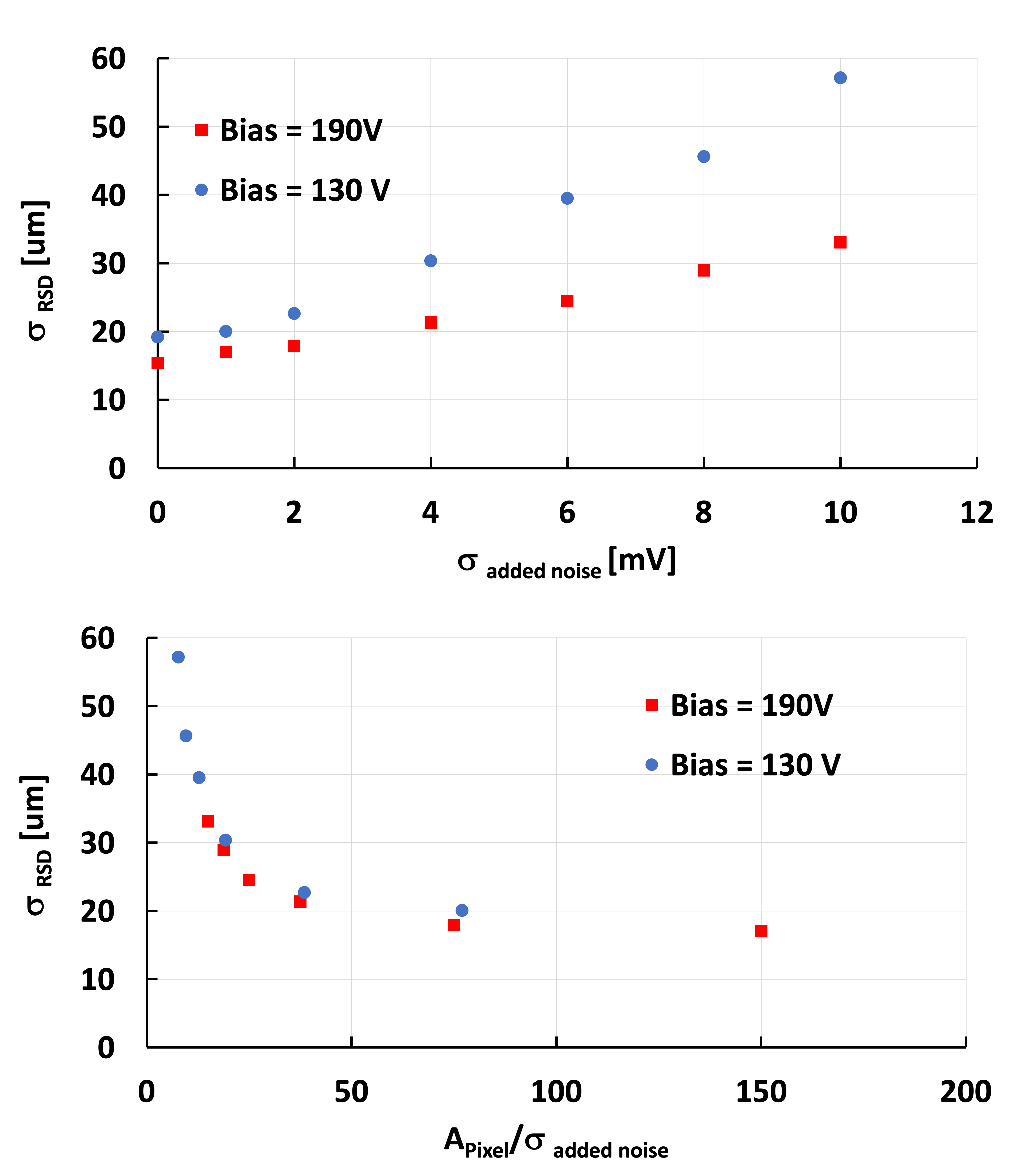}
\caption{Degradation of the spatial resolution as a function of added noise. Top: spatial resolution plotted versus the sigma of the added Gaussian noise. Bottom: the same data plotted versus the signal-to-added-noise ratio. }
\label{fig:AddedANoise}
\end{center}
\end{figure}

The effect of the electronic noise on the hit position has been studied by adding an uncorrelated Gaussian noise to each of the amplitudes $A_i$  used in the amplitude reconstruction. Figure~\ref{fig:AddedANoise} (top plot) reports the evolution of the spatial resolution for the dataset taken at 130 V and 190 V as a function of the added RMS noise, while  Figure~\ref{fig:AddedANoise} (bottom plot) shows the same data points against the signal-to-added-noise ratio  $A_{pixel}/\sigma_{added \; noise}$.  The degradation of the spatial resolution is rather mild as a function of the added noise, and it depends linearly on the noise-to-signal ratio. For values of  $A_{pixel}/\sigma_{added \; noise}$ above $\sim$ 50, the measured spatial resolution is reached.

\begin{figure}[htb]
\begin{center}
\includegraphics[width=0.8\textwidth]{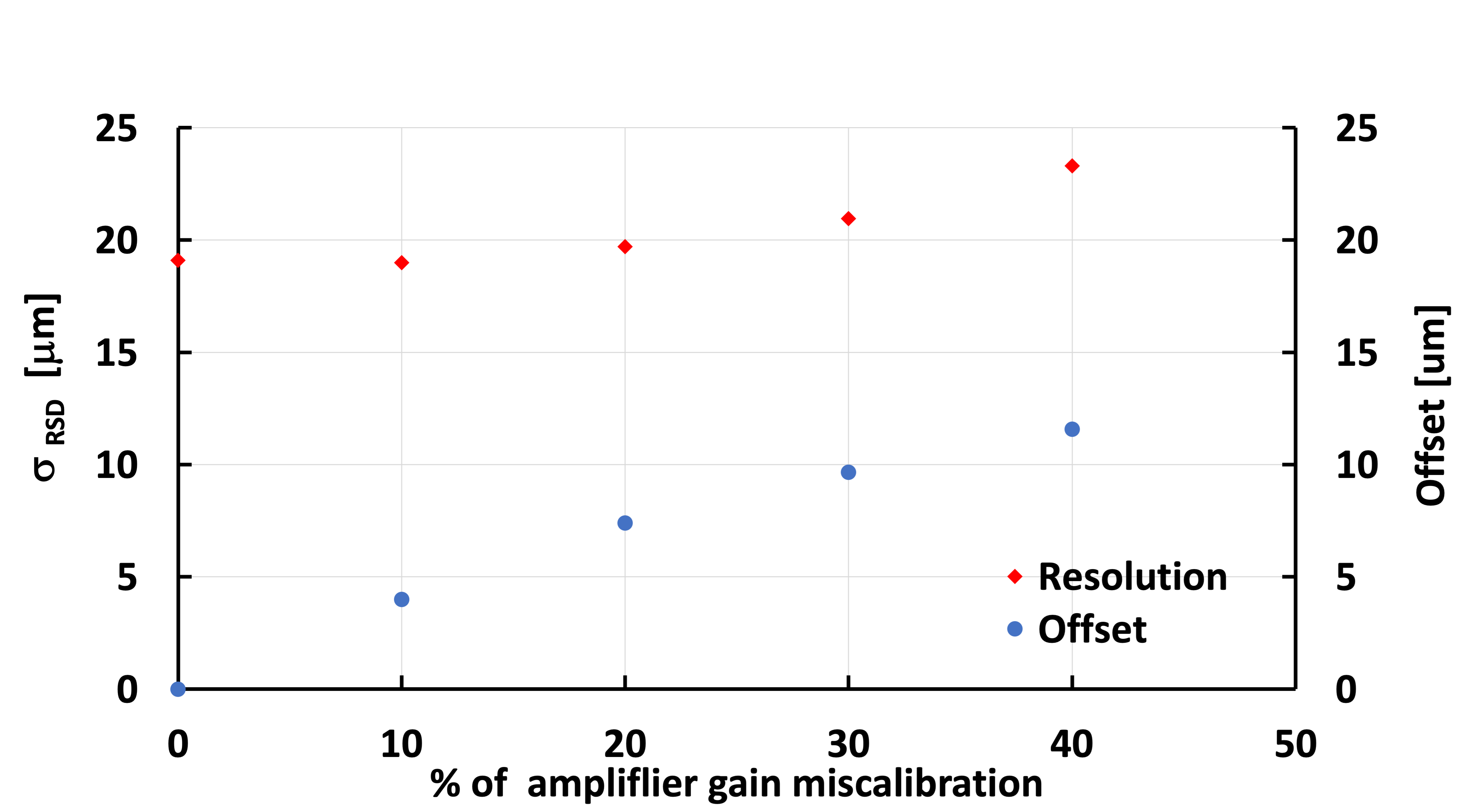}
\caption{Position resolution and offset as a function of the \% of one amplifier gain miscalibration. The study is performed using the run with Bias =  130 V.}
\label{fig:MisCal}
\end{center}
\end{figure}

The effects of incorrectly calibrating an amplifier are shown in Figure~\ref{fig:MisCal} for the run taken at  Bias = 130 V. The result is obtained by calculating the hit position resolution and offset while increasing up to 40\% the gain of one of the four amplifiers.   The result shows that the position resolution increases by about 20\%, and the difference between the mean RSD hit position and the tracker position goes from 0  to 12 \microns for a 40\% amplifier gain miscalibration.  

These two systematic studies demonstrate that the RSD spatial resolution remains very good even if the readout is much noisier than FAST2, an important consideration in view of using RSDs in much larger systems.

\clearpage

\subsection{Temporal resolution}
\begin{figure}[htb]
\begin{center}
\includegraphics[width=0.9\textwidth]{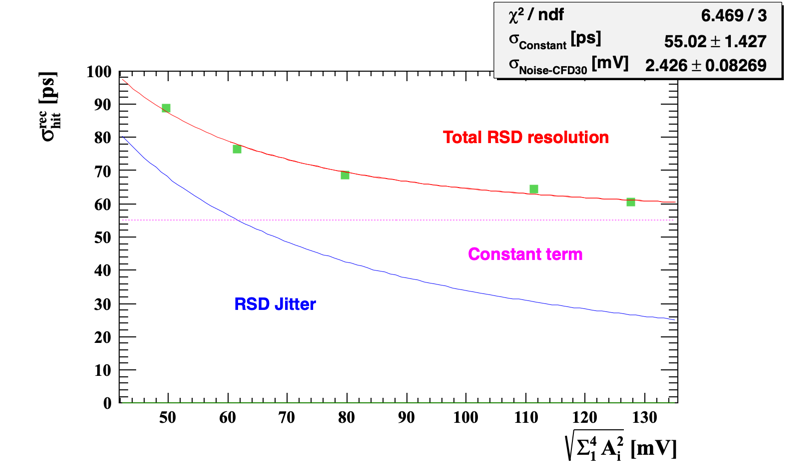}
\caption{Temporal resolution for the 3 pixels read out by FAST2 EVO1 channels.  The resolution is fitted as the sum of a jitter and a constant term.}
\label{fig:TEVO2}
\end{center}
\end{figure}

Given the difficulties in computing the inverse of the covariance matrix, the hit time $t_{hit}^{rec}$ has been calculated using Eq.~\ref{eq:tjitter}. 
The RSD temporal resolution $\sigma_{hit}^{rec}$ is calculated by subtracting in quadrature from the RMS of the distribution $t_{trigger}-t_{hit}^{rec}$ the resolution of the trigger, $\sigma_{trigger} $ = 12 ps. Since the EVO1 and EVO2 channels of FAST2 have a different signal-to-noise ratio, the results are based only on events collected by the three pixels fully read out by EVO1 channels (see Figure~\ref{fig:SR} for details).  The results are reported in Figure~\ref{fig:TEVO2} as a function of  $\sqrt{ \sum_i^4 A^2_i}$.  The best resolution obtained in this study is $\sigma_{hit}^{rec}$ =  60.6 ps, about 20 ps higher than the intrinsic RSD time resolution~\cite{tornago2020resistive}. The resolution, fitted as the sum in quadrature of a constant and a jitter term
\begin{equation}
\sigma_{hit}^{rec} = \sqrt{(\sigma^{time-constant})^2+ (\frac{\sigma^{CFD30}\times t^{rise} }{\Sigma_i^4 A^2_i})^2},
\end{equation}
  is dominated by the constant term.

\begin{figure}[htb]
\begin{center}
\includegraphics[width=1\textwidth]{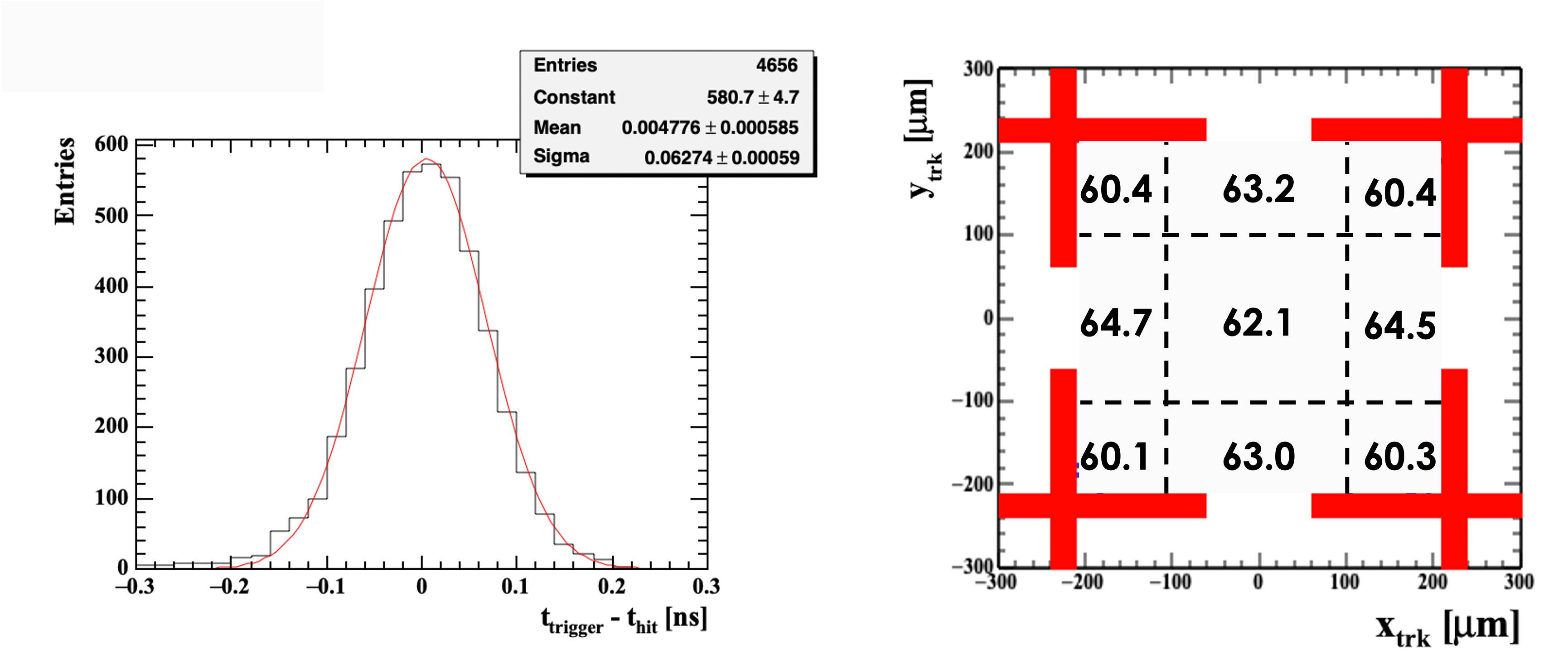}
\caption{Left: the distribution t$_{trigger}$ - t$^{rec}_{hit}$ (Bias = 200 V). Right: $\sigma_{hit}^{rec}$  as a function of the hit position in the pixel (Bias = 200 V).}
\label{fig:Tall}
\end{center}
\end{figure}

Figure~\ref{fig:Tall} (left) shows, for the highest gain point, the distribution of the difference t$_{trigger}$ - t$^{rec}_{hit}$ while Figure~\ref{fig:Tall} (right) shows $\sigma_{hit}^{rec} $ as a function of the hit position in the pixel. The temporal resolution is uniform over the pixel surface, indicating that the correction for signal delay is accurate. 

An insight into the origin of the constant term can be obtained by studying the correlated and uncorrelated parts of the temporal resolution of the 4 estimators $t_i^{rec}$. The expression of $\sigma^{rec}_i$, Eq.~\ref{eq:timeres}, can be rewritten as:  
\begin{equation}
\label{eq:timecor}
(\sigma^{rec}_i)^2 =   (\sigma^{cor})^2 + (\sigma^{uncor}_i)^2
\end{equation}

where $\sigma^{cor}$ is the part of the resolution common to all electrodes, and  $\sigma^{uncor}_i$ is the uncorrelated part. By selecting regions of the pixel equidistant from the i and j electrodes, the two terms $\sigma^{uncor}_i$ and  $\sigma^{uncor}_j$ become identical. With this selection, the RMS values of the distributions $t_i^{rec} - t_j^{rec}$ and  $t_i^{rec} - t_{trigger}$ can be written as:

\begin{flalign}
(\sigma^{rec}_{i,j})^2 & =  2 \times (\sigma^{uncor})^2, \\
(\sigma^{rec}_{i,trigger})^2 & =   (\sigma^{cor})^2 + (\sigma^{uncor}_i)^2+ \sigma_{trigger}^2.
\end{flalign}
This yields to
\begin{flalign}
\sigma^{uncor} & =  (\sigma^{rec}_{i,j})/\sqrt{2},\\
\sigma^{cor}  &= \sqrt{ (\sigma^{rec}_{i,trigger})^2- (\sigma^{uncor})^2- \sigma_{trigger}^2}.
\end{flalign}

\begin{figure}[htb]
\begin{center}
\includegraphics[width=.8\textwidth]{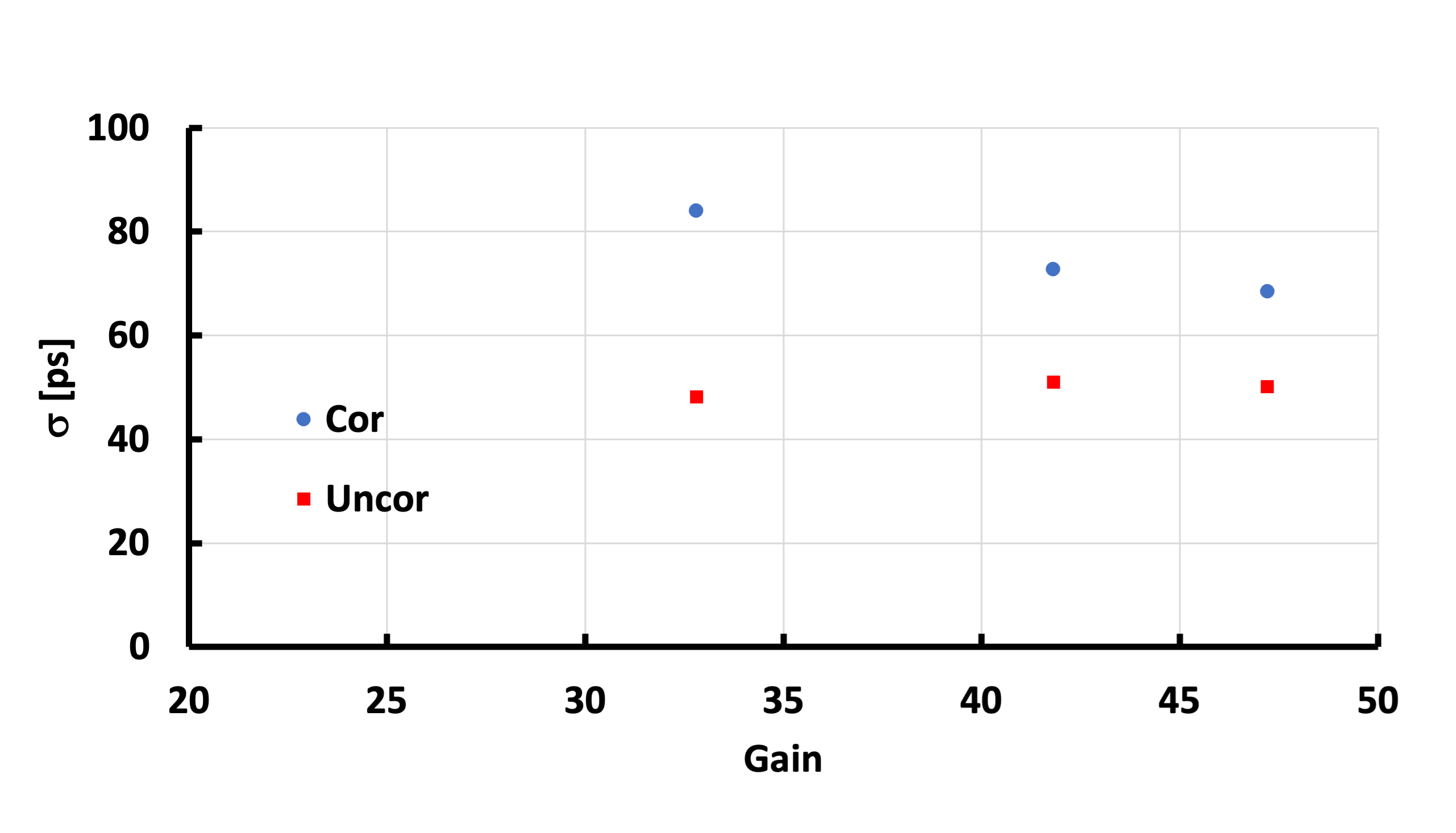}
\caption{Correlated and uncorrelated temporal resolution as a function of the sensor gain.}
\label{fig:CorUncor}
\end{center}
\end{figure}

The values of $\sigma^{uncor}$ and $\sigma^{cor}$ are reported as a function of gain in Figure~\ref{fig:CorUncor}. The uncorrelated part decreases with amplitude, indicating that is driven by the jitter. The correlated component is instead constant as a function of amplitude, determined by the Landau noise and the amplifier performance.

\subsection{Using delays to determine the hit position}

\begin{figure}[htb]
\begin{center}
\includegraphics[width=.8\textwidth]{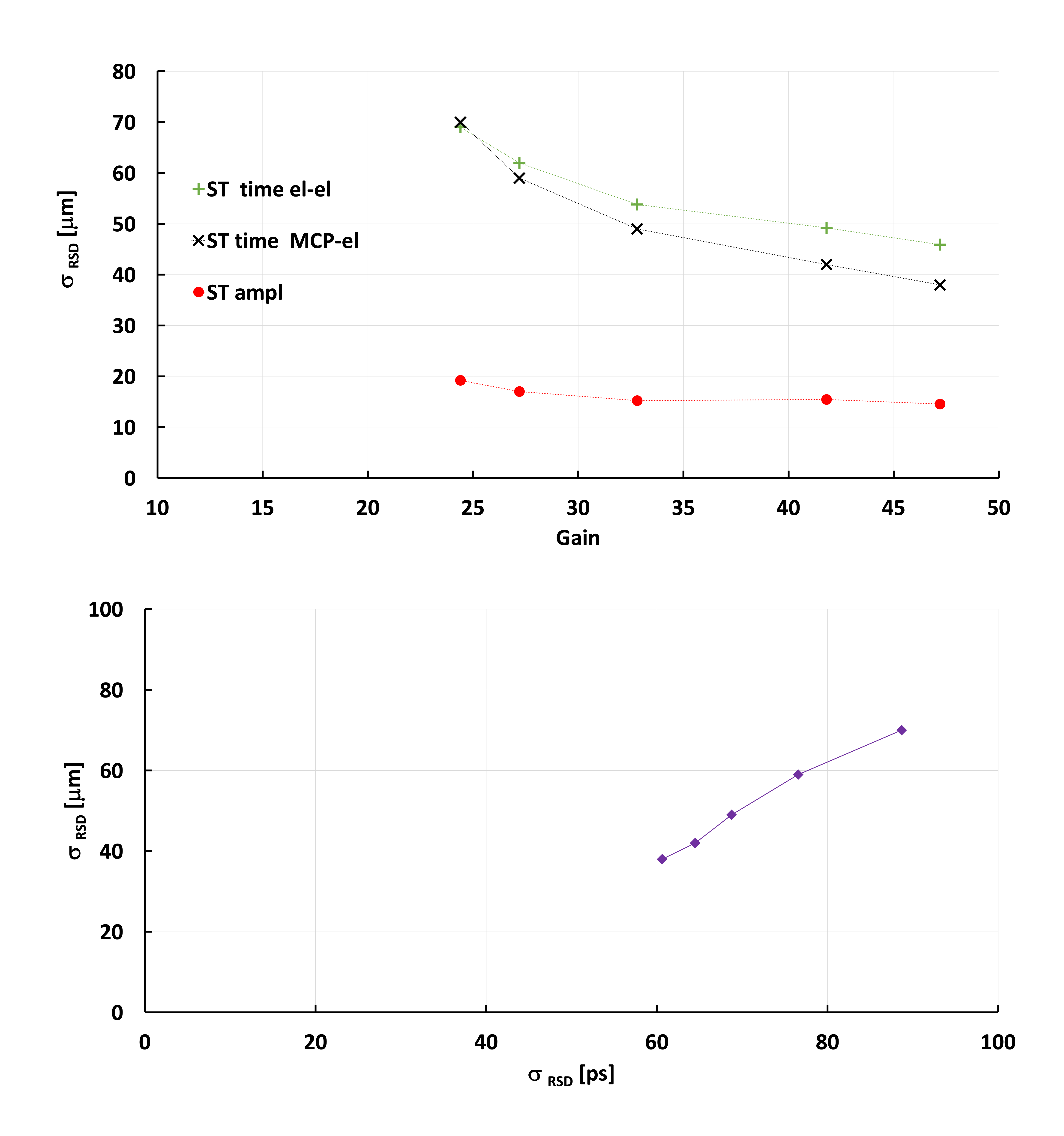}
\caption{Top: Position resolution as a function of the gain obtained with 3 different methods: ST$^{ampl}$, ST$^{time \; trigger-el}$, and  ST$^{time \;  el-el}$. Bottom: correlation between the temporal resolution and spatial   ST$^{time \;  trigger-el}$ resolution.}
\label{fig:STimedet}
\end{center}
\end{figure}

The signal delays between the hit point and each of the 4 electrodes can be used to calculate the hit position, following a procedure analogous to ST$^{ampl}$. Two  delay types can be used in the position determination: (i) the delays between each electrode and the trigger (ST$^{time \; trigger-el}$) and  (ii) between each pair of electrodes (ST$^{time \;  el-el}$). For each method, the corresponding templates were calculated using test beam data. The top part of Figure~\ref{fig:STimedet} shows the results of these studies and, for comparison,  the results of ST$^{ampl}$: the best resolution obtained with  ST$^{time \; trigger-el}$ is about $\sigma \sim $ 38 \microns,  approximately twice that of ST$^{ampl}$.  Figure~\ref{fig:STimedet} (bottom) shows the correlation between the ST$^{time \; trigger-el}$  spatial resolution and the temporal resolution: a temporal resolution of 40 ps would yield a spatial precision of 15 \micron, a result comparable to those obtained with ST$^{ampl}$.

\section{Conclusions}

This paper reports on the spatial and temporal resolutions of an RSD 450~µm pitch pixels array. The sensor matrix used in this study is part of the second FBK RSD production (RSD2),  and it consists of seven 450~\microns pitch pixels with cross-shaped electrodes, covering an area of about 1.5~mm$^2$. The electrodes were read out by the FAST2 ASIC, a 16-channel amplifier fully custom ASIC developed by INFN Torino using 110 nm CMOS technology. The study was performed at the DESY test beam facility with a 5~GeV/c electron beam. 
 Key findings include achieving a position resolution of $\sigma_{x}$ = 15 \microns, approximately 3.5\% of the pitch; standard pixelated sensors would require about 80 times more pixels to achieve similar spatial resolution. The temporal resolution is $\sigma_t$ = 60 ps, predominantly determined by the FAST2 resolution. The study also highlights the 100\% fill factor and homogeneous resolutions over the entire matrix surface achieved by RSD sensors. 
These results highlight the potential of RSD technology in applications requiring high spatial and temporal resolutions, offering a promising avenue for future developments in particle detection and imaging technologies.

\section*{Acknowledgments}

The measurements leading to these results have been performed at the test beam facility at DESY Hamburg (Germany), a member of the Helmholtz Association (HGF). This project has received funding from the European Union Horizon Europe research and innovation programme under grant agreement No 101057511.
We kindly acknowledge the following funding agencies and collaborations: RD50, INFN – FBK agreement on sensor production; Dipartimento di Eccellenza, Univ. of Torino (ex L. 232/2016, art. 1, cc. 314, 337); Ministero della Ricerca, Italia, PRIN 2017, Grant 2017L2XKTJ – 4DinSiDe; Ministero della Ricerca, Italia, FARE,  Grant R165xr8frt\_fare, Grant TRAPEZIO 2021 Fondazione San Paolo, Torino.  

\bibliography{NC_bibfile_2022_RSD2}

\end{document}